\documentclass[twocolumn]{aastex631}

\usepackage{xspace}
\newcommand{\g}{$\gamma$\xspace}
\newcommand{\hi}{H$\,${\sc i}\xspace}
\newcommand{\hii}{H$\,${\sc ii}\xspace}
\newcommand{\nh}{$N_{\rm{H}}$\xspace}
\newcommand{\nhi}{$N_{\text{H}\,\textsc{i}}$\xspace}

\newcommand{\ttau}{$\tau_{353}$\xspace}
\newcommand{\bb}{$\beta$\xspace}
\newcommand{\Iunits}{MJy sr$^{-1}$\xspace}
\newcommand{\minus}{\scalebox{0.75}[1.0]{$-$}}
\newcommand{\nhplus}{$N_{\text{H}\,\text{+}}$\xspace}

\graphicspath{{./}{Figures/}}

\begin{document}

\title{Evidence for large-scale excesses associated with low \hi column densities in the sky \\
 I. Dust excess
}

\author[0000-0003-1404-0524]{Jean-Marc Casandjian} 
\affiliation{Université Paris-Saclay, Université Paris Cité, CEA, CNRS, AIM, F-91191, Gif-sur-Yvette, France}
\author[0000-0002-8784-2977]{Jean Ballet} 
\affiliation{Université Paris-Saclay, Université Paris Cité, CEA, CNRS, AIM, F-91191, Gif-sur-Yvette, France}
\author[0000-0003-3274-674X]{Isabelle Grenier} 
\affiliation{Université Paris Cité, Université Paris-Saclay, CEA, CNRS, AIM, F-91191, Gif-sur-Yvette, France}
\author{Quentin Remy} 
\affiliation{Max-Planck-Institut fur Kernphysik, Saupfercheckweg 1, D-69117 Heidelberg, Germany}

\correspondingauthor{Jean-Marc Casandjian}
\email{casandjian@cea.fr}

\begin{abstract}
  Where dust and gas are uniformly mixed, atomic hydrogen can be traced through the detection of far-infrared (FIR) or UV emission of dust. We considered, for the origin of discrepancies observed between various direct and indirect tracers of gas outside the Galactic plane, possible corrections to the zero levels of the Planck-HFI detectors. We set the zero levels of the Planck High Frequency Instrument (HFI) skymaps as well as the 100 $\mu$m map from COBE/DIRBE and IRAS from the correlation between FIR emission and atomic hydrogen column density excluding regions of lowest gas column density. A modified blackbody model fit to those new zero-subtracted maps led to significantly different maps of the opacity spectral index $\beta$ and temperature $T$ and an overall increase in the optical depth at 353 GHz \ttau of 7.1$\times$10$^{-7}$ compared to the data release 2 Planck map. When comparing \ttau and the HI column density, we observed a uniform spatial distribution of the opacity outside regions with dark neutral gas and CO except in various large-scale regions of low \nhi that represent 25\% of the sky. In those regions, we observed an average dust column density 45\%  higher than predictions based on \nhi with a maximum of 250\% toward the Lockman Hole region. From the average opacity $\sigma_{e 353}$=(8.9$\pm$0.1)$\times$10$^{-27}$ cm$^2$  we deduced a dust-to-gas mass ratio of 0.53$\times$10$^{-2}$. We did not see evidence of dust associated to a Reynolds layer of ionized hydrogen. We measured a far-ultraviolet isotropic intensity of 137$\pm$15 photons s$^{-1}$cm$^{-2}$sr$^{-1}$$\AA$$^{-1}$ in agreement with extragalactic flux predictions and a near-ultraviolet isotropic intensity of 378$\pm$45 photons s$^{-1}$cm$^{-2}$sr$^{-1}$$\AA$$^{-1}$ corresponding to twice the predicted flux.
\end{abstract}




\section{Introduction} \label{sec:intro}

The hydrogen is the main component of the interstellar medium (ISM) gas. The neutral \hi, both in the cold and warm phases, is traced by its radio 21 cm line radiation, and its column density \nhi can be extracted from the 21 cm line radiation temperature often assuming a uniform spin temperature. While there is no direct large-scale observation of the warm ionized medium (WIM), its spatial distribution can be deduced from the density distribution of free electrons obtained from dispersion measures of pulsars with known distances. H$_2$ cannot be observed directly in its cold phase but should have the same distribution as its main collisional partner, the carbon-12 monoxide (CO), observed through its 2.6 mm J=1$\rightarrow$0 line. Interstellar gas is also detected indirectly by high-energy $\gamma$-ray emission produced by interactions of energetic cosmic rays (CR) with the gas nucleons mainly through the decay of secondary particles produced in hadron collisions. Once separated from other \g-ray contributors and deconvolved from CR intensity and production cross section, a complete census of interstellar hadrons can be achieved with, however, a worse angular resolution than radio and millimeter surveys.

The dust, mainly heated by far-ultraviolet (FUV) photons, radiates from far-infrared (FIR) to microwave frequencies. High-resolution observations at those frequencies provide its column density after disentangling from additional sky emissions and after correcting for dust temperature. Alternatively dust column density can be deduced from reddenings of background stars at optical wavelengths.

In the ISM collisions between dust grains and gas particles lead to a coupling between the gas and the dust. In the diffuse phases of the local ISM, we expect the dust spatial distribution to correlate with that of the gas. In principle, tracers of gas and tracers of dust should also spatially correlate.

While templates of dust column density mostly correlate with templates of \nh \citep{1988ApJ...330..964B, 2014A&A...566A..55P} discrepancies exist. In the outer envelopes of molecular clouds where the CO molecule is strongly affected by UV photodissociation, H$_{2}$ can exist without CO being present \citep{ 1988ApJ...334..771V, 2010ApJ...716.1191W, 2011MNRAS.412..337G}. This effect, in addition to optically thick \hi and nonuniformity of the spin temperature, lead to a large quantity of Dark Neutral Medium (DNM) gas not traced by H templates but observed through the combination of their dust and $\gamma$-ray emissions \citep{2005Sci...307.1292G, 2011A&A...536A..19P, 2018A&A...611A..51R}. Independently from DNM, toward dense CO clouds, the mantle accretion on the dust grains and the formation of grain aggregates enhance dust FIR emissivity \citep{ 2015A&A...579A..15K, 2017A&A...601A..78R}.

At low column density \nhi of a few 10$^{20}$ cm$^{-2}$, those issues are not relevant, and the spatial distribution of dust column density extracted from FIR observation should follow accurately that of \nhi. In \cite{2014A&A...571A..11P} (hereafter Planck XI), Figure 21 shows the correlation between those two templates. The uncertainty of each point being notably smaller than the error bar displaying the standard deviation in the \nhi bin, we observe a systematic departure from linearity for \nhi$>$2$\times$10$^{20}$ cm$^{-2}$. In this figure, for low \nhi from 2 to 3$\times 10^{20}$ cm$^{-2}$, \nhi and the dust column density are still correlated but with a larger proportionality coefficient (opacity) in addition to a strong offset. The departure from a simple linear scaling between \textbf{low} \nhi and FIR dust emission has been the subject of numerous studies emphasizing the linear correlation below \nhi of a few 10$^{20}$ cm$^{-2}$ and attributing the departure at intermediate column density to the hidden presence of H$_2$ or to dust overemission \citep{1996A&A...312..256B, 2014A&A...566A..55P, 2014ApJ...780...10L, 2014ApJ...783...17L, 2017ApJ...846...38L, 2020ApJ...899...15M}.

The larger value of the opacity (than the one quoted in Planck XI) observed for \nhi between 2 and 3$\times 10^{20}$ cm$^{-2}$ agrees with that needed to produce a DNM map in agreement with the Fermi Large Area Telescope (LAT) \g-ray diffuse observations (Figure 4, top of \cite{2016ApJS..223...26A}). Increasing the dust opacity implies lowering the zero levels to remain consistent with Planck data. This has led us to investigate a potential issue with the zero levels of the Planck detectors. Planck zero levels (offsets + cosmic infrared background, hereafter CIB) were calculated as the y-intercept of the linear correlation between dust FIR and \nhi at lowest \nhi (Planck XI).
It assumes that only a small amount of dust exists apart from the one associated with \hi. \cite{1996A&A...312..256B} and \cite{1998ApJ...508...74A} validated this hypothesis. \cite{1998ApJ...508...74A} excluded any contribution from dust associated to H$_2$ and estimated a maximum of 20\% for the contribution of \hii traced by pulsar dispersion measure and H$_{\alpha}$ observations. Still, the ionized gas column density that cannot be measured on large scales is uncertain. In addition one cannot exclude that the dust emissivity could increase in very diffuse \hi.

In this work, we therefore derived the zero level of the Planck ﻿High Frequency Instrument (HFI) at 857 GHz outside the regions of lowest \nhi and in a wide region where the \g-ray data is consistent with the total gas column densities being the observed \nhi values. Using the new zero levels, we recalculated the dust optical depth at 353 GHz \ttau and we compared it to \nhi. Since dust also scatters UV photons, we verified that the new optical depth agrees with FUV and near-ultraviolet (NUV) observations.

\section{Data}

\subsection{Planck-HFI}

We downloaded the Planck-HFI skymaps from the Planck Legacy Archive\footnote{\url{http://pla.esac.esa.int/pla/\#home}}. We selected the 2015 data release 2 (PR2) for a better comparison with the work of \cite{2016A&A...596A.109P} (hereafter Planck XLVIII). No major difference was observed with the release 3 version. We made use of the generalized needlet internal linear combination (GNILC) products (Planck XLVIII) voided of CIB anisotropy and recommended for dust studies. As similarly done in Planck XI and in Planck XLVIII, we utilized the 3 maps associated to the highest HFI frequencies: 353, 545, and 857 GHz\footnote{COM\_CompMap\_Dust-GNILC-F353\_2048\_R2.00.fits, COM\_CompMap\_Dust-GNILC-F545\_2048\_R2.00.fits, COM\_CompMap\_Dust-GNILC-F857\_2048\_R2.00.fits }. We did not measure any significant improvement when adding to the fit lower-frequency maps. Following Planck XLVIII, we also used half-mission maps to derive the instrumental noise. 

Since HFI spectral responses are not uniform, we must account for the intensity variation within a detector bandpass due to the signal spectral profile. We obtained those color-correction coefficients from dedicated software routines provided by the Planck collaboration\footnote{\url{https://wiki.cosmos.esa.int/planck-legacy-archive/index.php/Unit_conversion_and_Color_correction}}.

We transformed all HFI-Planck intensity maps as well as other skymaps used in this work into HEALPix standard \citep{2005ApJ...622..759G}  with a resolution parameter $N_{side}=512$ corresponding to mean pixel spacing of 0$^{\circ}$.11.

\subsection{100 $\mu$m}

\begin{figure}
  \centering
  \includegraphics[width=\hsize]{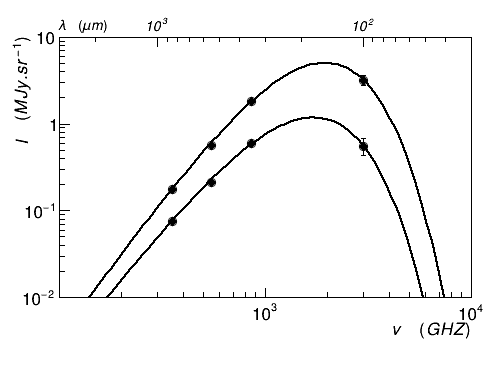}
  \caption{Thermal spectrum of a pixel located in the Lockman Hole (lower curve) and at $l$=45$^{\circ}$ and $b$=30$^{\circ}$ (higher curve). We show the intensity at 100 $\mu$m and 857, 545, 353 GHz and associated uncertainties together with the best modified blackbody fit (black lines).}
  \label{fig_thermal_spectra_one_pixel}
\end{figure}

100 $\mu$m is a crucial wavelength for the fit of a modified blackbody (MBB) spectrum. The associated skymap intensity corresponds for T$\sim$20 K to the only point in the spectral fit not located on the same side of the modified Planck curve as the Planck-HFI frequencies (Figure \ref{fig_thermal_spectra_one_pixel}). Its value mostly drives the temperature parameter. Based on Planck XI and Planck XLVIII we choose to use the hybrid map\footnote{IRIS\_combined\_SFD\_really\_nohole\_nosource\_4\_2048.fits}  combining at large scale the map of \cite{1998ApJ...500..525S} (itself a composite of COBE/DIRBE and IRAS) and at smaller scale the IRIS version of IRAS \citep{2005ApJS..157..302M}. We downloaded this map from the Planck Legacy Archive website. We approximated the color correction of this composite map with coefficients given by the COBE/DIRBE collaboration\footnote{\url{https://lambda.gsfc.nasa.gov/data/cobe/dirbe/ancil/colcorr/DIRBE_COLOR_CORRECTION_TABLES.ASC}\label{refnote}}.

\subsection{DIRBE}

While we did not use the COBE/DIRBE maps for the MBB fit because of their low spatial resolution, we used them for calibration purposes. We utilized the maps at 140 and 240 $\mu$m\footnote{\url{https://lambda.gsfc.nasa.gov/product/cobe/dirbe_zsma_data_get.cfm}}. Those maps, as the previous ones, have the zodiacal light and point sources removed; the intensity was color-corrected with the dedicated coefficients\textsuperscript{\ref{refnote}}.

\subsection{\nh} 

As we mentioned, under certain conditions, the dust optical depth derived from far-infrared surveys can be a good tracer for the total hydrogen column density \nh. We derived the atomic hydrogen column density \nhi from the HI4PI survey \citep{2016A&A...594A.116H} of brightness temperature ($T_\textrm{B}$) assuming an optically thin limit: $N_{\text{H}\,\textsc{i}}(l,b)=1.83\times10^{18}\,\text{cm}^{-2}\int T_\textrm{B}(l,b,v)dv$. In this study, we masked pixels with DNM and verified that this mask also covers all the molecular clouds.
The Doppler shifts of the \hi lines have provided the kinematic Galactocentric distances of the atomic and molecular gas.
We used this information to partition the gas column density map into several components as in \cite{2017A&A...601A..78R} and separated 11 Galactocentric annuli maps. We used this partition only for an accurate visualization of the low-latitude regions of our residual maps, not in the derivation of the zero levels or of the MBB parameters.
We removed high velocity clouds from the maps. 
Finally, to calculate the residuals between dust and gas, we corrected the modeled gas distribution for the Galactic metallicity gradient which affects the dust abundance. We used the average abundance of O and Si \citep{2007A&A...462..943C,2015A&A...580A.127K} normalized to 1 for the solar neighborhood.

\section{Initial Estimates of the Zero Levels of FIR maps}

\subsection{Method}

In this section, we describe our method to obtain initial estimates of the zero levels of the GNILC dust intensity skymaps used in this work.
 The zero levels include both the intrinsic detector offsets and the spatially uniform CIB monopole.

 Considering that contamination of the dust thermal emission by free-free emission decreases at high frequencies and that the 857 GHz intensity better correlates with gas column densities than the 100 $\mu$m intensity (see Figure 5 of Planck XI), we have first correlated $I_{857\rm{GHz}}$ and \nhi in a mask that we called \emph{Pegasus-Aquarius} corresponding to the region of longitude 0$^{\circ}$ to 180$^{\circ}$ and of latitude $\minus$60$^{\circ}$ to $\minus$15$^{\circ}$. Building upon our experience with Fermi-LAT interstellar modeling \citep{ 2015ApJ...806..240C,  2016ApJS..223...26A} for which we observed a good agreement between diffuse \g-rays and a model based on \nhi or \ttau in this region, the FIR intensity of dust and \nhi column density are linearly correlated in this region.
We excluded the regions with potential contamination by H$_2$ or significant \hi self-absorption (DNM). For that we compared the \hi column density provided by the dust: $\tau_{353}/\sigma_{e 353}$ to \nhi from 21 cm survey and masked regions where $\tau_{353}/\sigma_{e 353}-N_{\text{H}\,\textsc{i}}>10^{20}$ cm$^{-2}$. To remove a possible bias related to the choice of the initial value for the opacity $\sigma_{e 353}$ and \ttau, we strongly reduced to $N_{side}$=8 the resolution of both \ttau and \nhi in the calculation of the exclusion region. We proceeded iteratively starting from an opacity $\sigma_{e 353}$=6.3$\times$10$^{-27}$ cm$^2$ given by Planck XI and optical depth \ttau obtained in Planck XLVIII. We note that, while we used a lower angular resolution to estimate the extent of the region with possible H$_2$ or self-absorbed \hi contamination, the resolution of the mask itself was kept at $N_{side}$=512 for the correlation studies. We show the Pegasus-Aquarius mask in the inset of Figure \ref{fig_zero_857} (later).

  We derived the zero levels at 545, 353 GHz, and 100 $\mu$m from a correlation of these intensities with $I_{857\rm{GHz}}$ in a dedicated mask called \emph{Isothermal}. In this mask, we selected pixels for which the MBB parameters $T$ and \bb did not depart by more than half a standard deviation from the sky average values. This ensures a selection of pixels with rather uniform spectral parameters; it also reduces potential inter-calibration problems between the various FIR telescopes used in this work. Again we proceeded iteratively starting from the maps of \bb and $T$ from Planck XLVIII. To exclude pixels with very low or high \nhi, we required that the mask pixels have \nhi within 2-15$\times 10^{20}$ cm$^{-2}$. We also excluded pixels with an absolute ecliptic latitude smaller than 15$^{\circ}$ to avoid zodiacal light contamination. We show the Isothermal mask in the inset of Figure \ref{fig_zero_545_353} (left). For the zero level of the 100 $\mu$m map, we also used an alternative method in which we correlated directly $I_{100\rm{\mu}}$ with \nhi. Since the correlation is poorer in the Pegasus-Aquarius mask due to thermal changes in the dust across this region (see Figure \ref{fig_tau_T_beta} later), we performed the correlation in the Isothermal mask that reduces the effect of spectral variations. 

As we describe in Section \ref{sec_procedure}, we have used these initial zero levels to fit an MBB to the intensities observed at the four frequencies in each sky pixel. The final set of zero levels was derived from the monopoles of the residual maps that compared the observations and the best-fit MBB spectra. We measured the monopole of the residuals for pixels inside a mask called \emph{Monopole} for which $\tau_{353}/\sigma_{e 353}-N_{\text{H}\,\textsc{i}}<2\times10^{20}$ cm$^{-2}$ to exclude potential contamination by H$_2$ or significant \hi self-absorption and for pixels with an absolute ecliptic latitude larger than 15$^{\circ}$. We show the Monopole mask in the inset of Figure \ref{fig_calibration} (top, left). 

\subsection{Initial Zero Level at 857 GHz}

We first ensured that the correlation between $I_{857\rm{GHz}}$ and \nhi is not significantly biased by the potential presence of ionized hydrogen in regions encompassed by the Pegasus-Aquarius mask. To do so, we fitted the $I_{857\rm{GHz}}$ map with a linear combination of \nhi, isotropic, and \nhplus maps. We used two possible spatial distributions for \nhplus column densities: one from the electron column density ($I_{NE2001}$) predicted by NE2001 \citep{2002astro.ph..7156C} and one from the H$_{\alpha}$ skymap intensity ($I_{H_{\alpha}}$) \citep{2010ASPC..438..388H}. The fit did not detect any contribution to the $I_{857\rm{GHz}}$ intensity from these \nhplus distributions. It set their contribution to zero with an upper limit of 5$\times 10^{16}$ cm$^{-2}$ for potential \nhplus contamination in this region according to the NE2001 map.

In the Pegasus-Aquarius mask, \nhi column densities range from 2 to 5$\times 10^{20}$ cm$^{-2}$. In this regime, the gas is little shielded from UV radiation, and H$_2$ is efficiently photodissociated \citep{2020A&A...643A..36B,2010ApJ...716.1191W,2006ARA&A..44..367S}. Based on the scaling law between FUV fluorescent H$_{2}$ emission and FIR dust emission (Figure 14c of \cite{2017ApJS..231...21J}) we estimated that the average N(H$_2$) column density in the mask region is only 2$\times 10^{17}$ cm$^{-2}$. This value is consistent with the 5$\times 10^{17}$ cm$^{-2}$ average N(H$_2$) column density inferred from the Far Ultraviolet Spectroscopic Explorer (FUSE) satellite observations toward Mrk 0509, Mrk 1513, NGC 1068 \citep{2006ApJ...636..891G}, and WD 0004+330 \citep{2003ApJ...595..858L}, which are all seen inside the Pegasus-Aquarius mask. We also stress that regions with column densities of CO-dark H$_2$ typically larger than $\sim$1$\times 10^{20}$ cm$^{-2}$ have been masked out in the construction of the Pegasus-Aquarius mask. By combining diffuse $\gamma$-ray data with dust optical depths (inferred from dust reddening or dust thermal emission), we have the means to detect optically thick \hi or diffuse H$_2$ column densities that are not accounted for by \hi and CO line data at the transition between the atomic and molecular phases \citep{2018A&A...616A..71R}. The N(H$_2$) column densities inferred in this CO-dark neutral interface match those seen by FUSE or those deduced from HCO+ observations \citep{2019A&A...627A..95L}. These dark-gas-rich regions have been masked out so that they do not bias the $I_{857\rm{GHz}}$ and \nhi correlation, hence the zero-point derivation.

\begin{figure}
  \centering
  \includegraphics[width=\hsize]{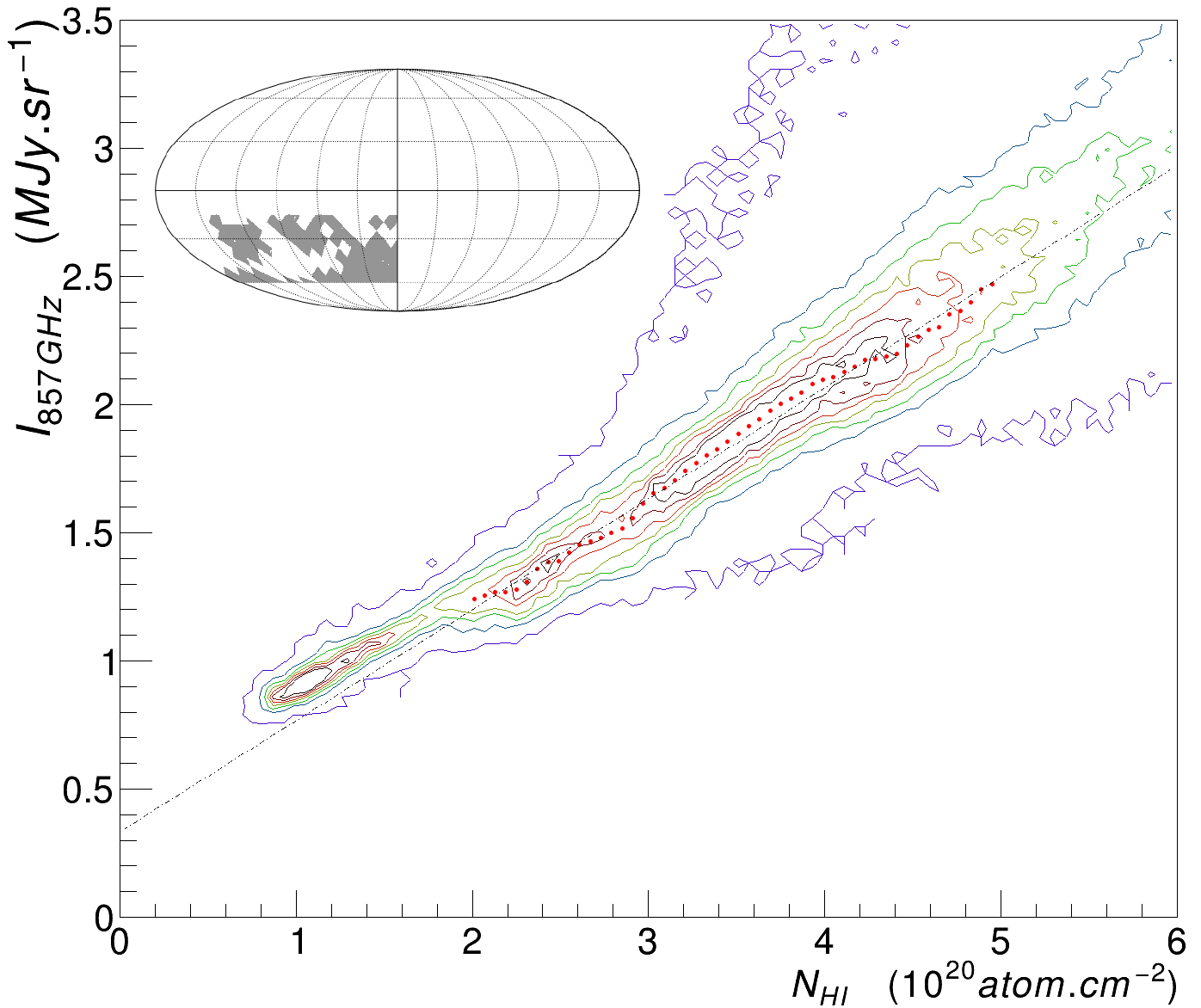}
  \caption{Correlation of $I_{857\rm{GHz}}$ and \nhi displayed with 7 contours linearly spaced between 1 and 250 pixel counts. We masked pixels outside the Pegasus-Aquarius mask shown in the inset. The red points are the centroids of the distribution in bins of \nhi for \nhi above the break at 2$\times 10^{20}$ cm$^{-2}$ and below 5$\times 10^{20}$ cm$^{-2}$ to ensure that the hydrogen is optically thin. We calculated the corresponding error bars, barely visible, by dividing the standard deviation in $I_{857\rm{GHz}}$ by the square root of the number of pixels in each \nhi bin. The dashed line corresponds to the best affine fit of those points. The y-intercept of the line represents to first order the zero level of $I_{857\rm{GHz}}$. We see that using $N_{\text{H}\,\textsc{i}}$$=$2$\times10^{20}$ cm$^{-2}$ as upper limit for the affine fit like in Planck XI and Planck XLVIII would lead to a zero value significantly larger.}
  \label{fig_zero_857}
\end{figure}

In Figure \ref{fig_zero_857} we represent the contours of the scatter plot of $I_{857\rm{GHz}}$ and \nhi in the Pegasus-Aquarius mask. We observe a very good affine correlation for \nhi larger than about 2$\times 10^{20}$ cm$^{-2}$; below this limit, we observe a break in the slope. We observe a similar break at a similar column density in Figure 21 of Planck XI. We derived and plotted the centroid position in each \nhi bin with a Gaussian curve fit. For \nhi larger than about 3$\times 10^{20}$ cm$^{-2}$ we observe an overemission of dust compared to what one would expect from hydrogen column density, which signals the residual presence of gas not traced by optically thin \hi in the mask. As a consequence, there is a tail at high $I_{857\rm{GHz}}$ in most of \nhi bins. To avoid any bias in the centroids' position, we restricted the upper limit of the Gaussian fit to a value corresponding to 1.05 times the centroid position. We fitted the centroid points for \nhi between 2 and 5$\times 10^{20}$ cm$^{-2}$, with a line whose y-intercept was $I_{0(857\rm{GHz})}$=0.33$\pm$0.03 \Iunits which represents the first-order zero for the 857 GHz Planck-HFI. Here the uncertainty includes systematic errors through a procedure detailed in Section \ref{systematic_uncertainties}.

In order to color-correct the HFI maps for $T$ and \bb, we repeated this operation three times, a first time with $T$ and \bb given by Planck XLVIII, then using the resulting MBB fit parameters for a next iteration.
As already mentioned we obtained the final value of this zero level in a second step described in Section \ref{sec_procedure}.

\subsection{Initial Zero Levels at 545 and 353 GHz}

\begin{figure}
  \centering
  \includegraphics[width=\hsize]{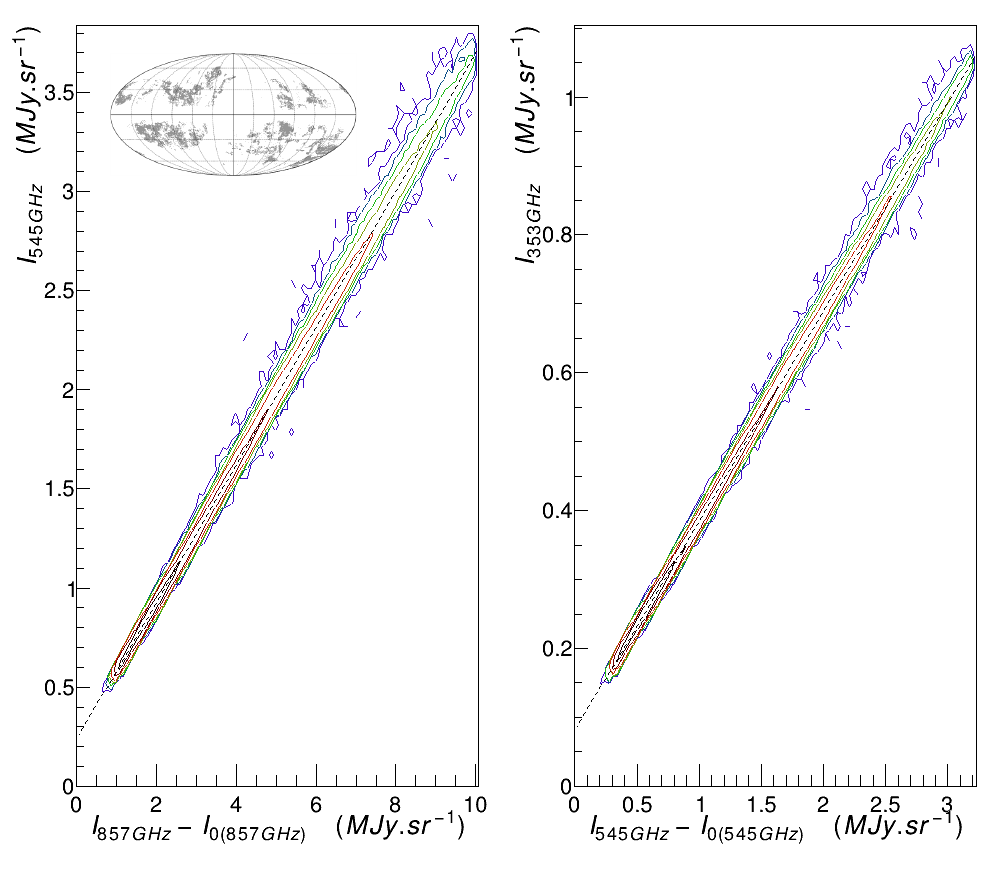}
  \caption{Correlation of $I_{545\rm{GHz}}$ and $I_{857\rm{GHz}}$ (left) and  $I_{353\rm{GHz}}$ and $I_{545\rm{GHz}}$ (right) displayed with 7 contours spaced in log between 1 and 5000 pixel counts. The dashed line corresponds to the best affine fit of the histogram points between the 5th and the 95th percentile of the dust intensity distribution (1.2 and 6.7 \Iunits for the left plot, and 0.4 and 2.2 \Iunits for the right plot). We masked data associated with the white pixels of the Isothermal mask shown in the inset.}
  \label{fig_zero_545_353}
\end{figure}

Once the initial zero level at 857 GHz was set, it was straightforward to derive the initial zero-level value for 545 and 353 GHz maps. We correlated $I_{857\rm{GHz}}$ and $I_{545\rm{GHz}}$ (Figure \ref{fig_zero_545_353} left), removed the zero level of $I_{545\rm{GHz}}$ and then correlated $I_{545\rm{GHz}}$ and $I_{353\rm{GHz}}$ (Figure \ref{fig_zero_545_353} right) in the Isothermal mask.
We obtained a very good correlation, and extracted the values of $I_{0(545\rm{GHz})}$=0.24$\pm$0.01 \Iunits and $I_{0(353\rm{GHz})}$=0.081$\pm$0.003 \Iunits for the initial zero levels at 545 and 353 GHz respectively from the y-intercept of an affine line fit.
Details about the calculation of the uncertainties are given in Section \ref{systematic_uncertainties}.
We have checked that the zero levels found in the Pegasus-Aquarius and in the Isothermal masks are fully compatible (within 1$\sigma$). For the rest of the analysis, we have chosen the Isothermal values that led to smaller uncertainties because of the larger number of pixels in the correlations.
The final zero values are calculated in a second step with the MBB fit residuals in Section \ref{sec_procedure}. We applied the same method for color correction as for $I_{857\rm{GHz}}$.

\subsection{Initial Zero Level at 100 $\mu$m}

\begin{figure}
  \centering
  \includegraphics[width=\hsize]{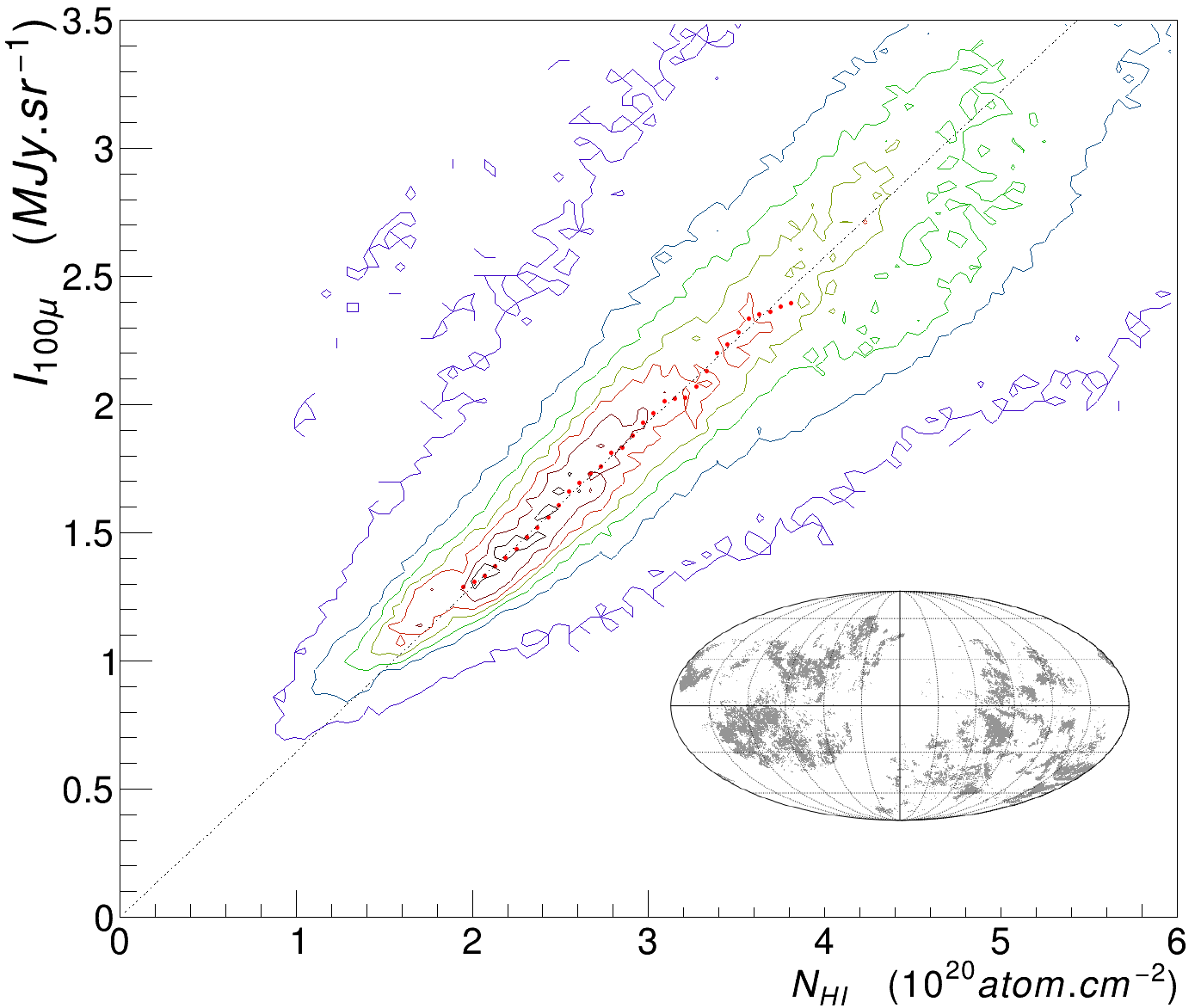}
  \caption{Correlation of $I_{100\rm{\mu}}$ and \nhi displayed with 7 contours linearly spaced between 1 and 250 pixel counts. The red points are the centroids of the distribution in bins of \nh. We calculated the corresponding error bars, barely visible, by dividing the standard deviation in $I_{100\rm{\mu}}$ by the square root of the number of pixels in each \nhi bin. The dashed line corresponds to the best affine fit of those points between 2 and 3.5$\times10^{20}$ cm$^{-2}$. We masked data associated with the white pixels of the Isothermal mask displayed in the inset that we extended in this graph to lower \nhi for display purpose.}
  \label{fig_zero_100}
\end{figure}

We first measured the initial zero level of the 100 $\mu$m intensity map from a simple correlation with \nhi in the optimized Isothermal mask. We show the resulting plot in Figure \ref{fig_zero_100} in which we extended \nhi to lower column density for display purpose. As we can see, the correlation is not as strong as at 857 GHz and becomes nonlinear for $N_{\text{H}\,\textsc{i}}$$\gtrsim$4$\times10^{20}$ cm$^{-2}$. Indeed, at 100 $\mu$m for $T\sim$20 K the intensity of the dust emission is not in the Rayleigh-Jeans regime but in the exponential decrease of the MBB spectrum (Figure \ref{fig_thermal_spectra_one_pixel}). Small variations of T (and to a lesser extent of \bb) lead to large variations of $I_{100\rm{\mu}}$ even for pixels within our Isothermal mask for which this nuisance is minimized. We found from an affine fit a y-intercept of -0.01$\pm$0.04 \Iunits.

\begin{figure}
  \centering
  \includegraphics[width=\hsize]{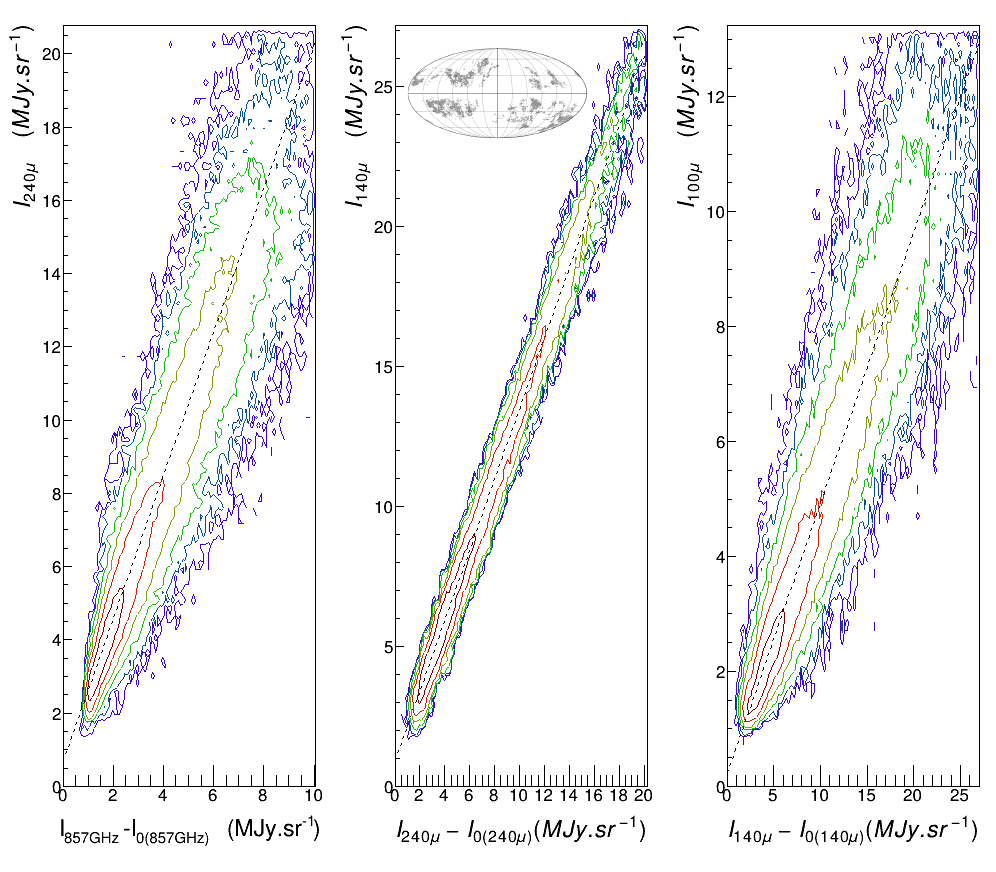}
  \caption{Correlation of $I_{240\rm{\mu}}$ and $I_{857\rm{GHz}}$ (left), $I_{140\rm{\mu}}$ and $I_{240\rm{\mu}s}$ (middle), and $I_{100\rm{\mu}}$ and $I_{140\rm{\mu}}$ (right) displayed with 7 contours spaced in log between 1 and 5000 pixel counts. The dashed line corresponds to the best affine fit of the histogram points between the 5th and the 95th percentile of the dust intensity distribution (1.2 and 6.7 \Iunits for the left plot, 2.3 and 13.5 \Iunits for the middle plot, and 2.8 and 18.0 \Iunits for the right plot). We masked data associated with the white pixels of the Isothermal mask shown in the inset.}
  \label{fig_zero_100_with_Dirbe}
\end{figure}

We used a second method to get the initial zero level at 100 $\mu$m in which we correlated in the same Isothermal mask $I_{857\rm{GHz}}$ and $I_{100\rm{\mu}}$ using as intermediate wavelength the COBE/DIRBE maps at 240 and 140 $\mu$m. We first correlated $I_{240\rm{\mu}}$ with $I_{857\rm{GHz}}$ (Figure \ref{fig_zero_100_with_Dirbe}, left), then $I_{140\rm{\mu}}$ with $I_{240\rm{\mu}}$ (Figure \ref{fig_zero_100_with_Dirbe}, middle), and finally $I_{100\rm{\mu}}$ with $I_{140\rm{\mu}}$ (Figure \ref{fig_zero_100_with_Dirbe}, right). At each step, we subtracted the zero level of the reference map. After an affine fit, we obtained $I_{0(240\mu)}$=0.7$\pm$0.1, $I_{0(140\mu)}$=0.9$\pm$0.1 and $I_{0(100\mu)}$=0.22$\pm$0.06 \Iunits respectively for the 240, 140, and 100 $\mu$m y-intercepts. As stated above, intensity maps were color-corrected iteratively. We note that since the COBE/DIRBE is absolutely calibrated, the value of the y-intercept at 240 $\mu$m and 140 $\mu$m should be close to the CIB. We will discuss this aspect in a forthcoming publication. Details about the calculation of the uncertainties are given in Section \ref{systematic_uncertainties}.

For the initial zero level of $I_{100\rm{\mu}}$ we therefore obtained $\minus$0.01$\pm$0.04 and 0.22$\pm$0.06 \Iunits. The former result is biased by strong effects of $T$ and \bb variations and the latter by propagation of errors from intermediate 240 and 140 $\mu$m zero levels.
Since this value is an important parameter for the MBB fit, especially for the temperature, we decided to use as initial zero level for $I_{100\rm{\mu}}$ a number randomly sampled between $\minus$0.1 and 0.2 \Iunits (see Section \ref{systematic_uncertainties}). A value larger than 0.2 would lead to a large quantity of negative pixels in the 100 $\mu$m map and a zero-level smaller than $\minus$0.1 lead to a suspicious increase in $T$ at the Galactic poles. For each realization, we derived the final zero levels in a second step from the MBB fit residuals.

\section{Modified Blackbody Fit}

\subsection{Procedure} \label{sec_procedure}

In this work we modeled the dust with an MBB spectrum that we fitted to observations at 100 $\mu$m, 857, 545 and 353 GHz with their zero offsets subtracted.
 For each pixel of the sky we fitted a parameterized MBB function $I_{\nu}(\tau_{353},T,\beta) = \tau_{353}B_{\nu}(T)(\nu/353 \,\text{GHz})^\beta$ where $B_{\nu}$ is the Planck function and where the parameters \ttau, $T$ and \bb were free to vary. 

For a correct color correction of the intensity maps we must assume an initial value for $T$ and \bb. We used the values given in Planck XLVIII, and iterated with initial values taken from the previous iteration. Three iterations were required after which we did not observe any significant improvement beyond 1$\sigma$. We verified that no major difference on $T$ and \bb was found when starting with a spatially uniform values $T$=20 K and \bb=1.4.
The fit procedure includes 2 steps. In a first step we fitted the MBB model to intensity maps with initial zero levels set from the correlation method described above. For each pixel we extracted the parameters \ttau, $T$ and \bb. Using those parameters we calculated MBB model predictions for the 4 frequencies and then calculated the residual maps of Planck-HFI and COBE/DIRBE intensities (without subtracting the preliminary zero levels) minus MBB models. We measured the monopole of the residuals for pixels inside the Monopole mask. We used those monopoles as final zero levels. 

\begin{figure}
  \centering
  \includegraphics[width=\hsize]{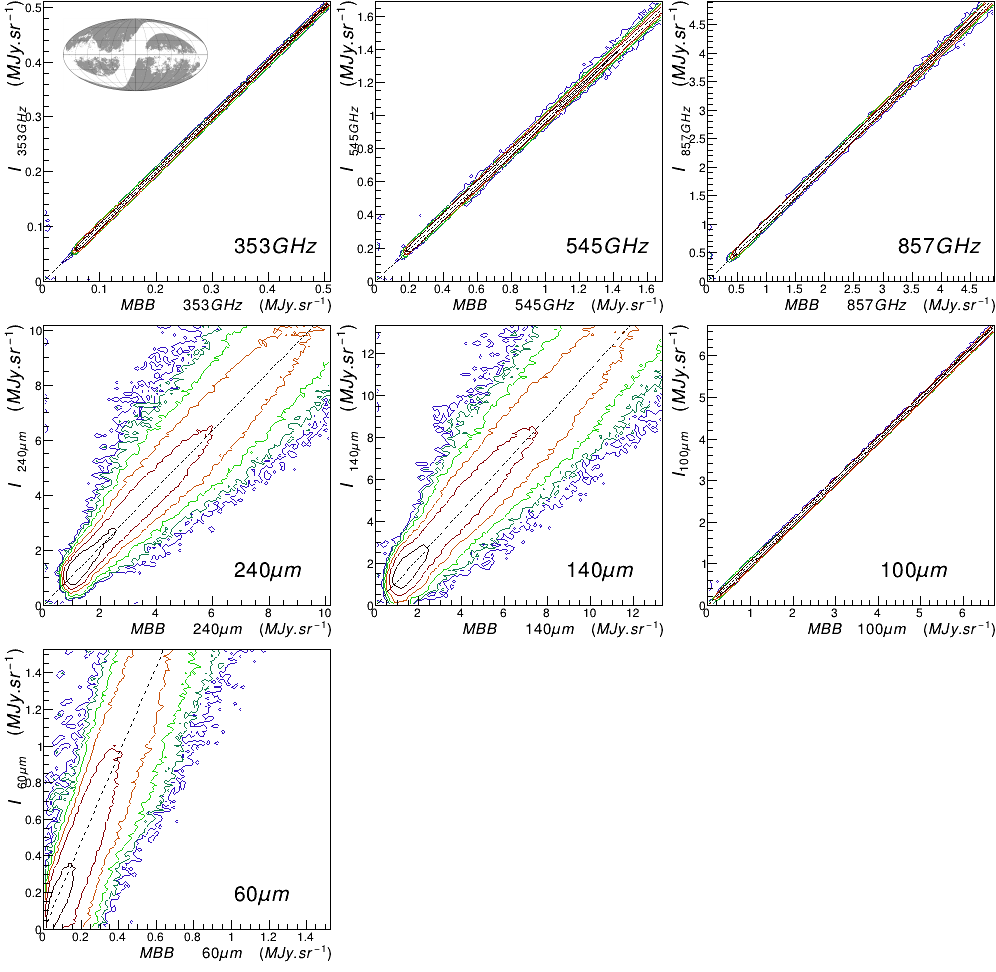}
  \caption{Correlation between intensity from observations and MBB predictions displayed with 8 contours spaced in log between 1 and 50,000 pixel counts. The gains given in Table \ref{table_zeros} correspond to the slope of the fitted black dashed line. We have subtracted the zero levels from observed intensities. We masked pixels with $\tau_{353}/\sigma_{e 353}-N_{\text{H}\,\textsc{i}}>2\times10^{20}$ cm$^{-2}$ and an absolute ecliptic latitude smaller than 15$^{\circ}$. We show the corresponding mask in the inset of the first plot.}
  \label{fig_calibration}
\end{figure}

In a second step, using those zero levels, we fitted again the MBB model to Planck-HFI and IRAS/DIRBE intensity maps and extracted the final values for \ttau, $T$ and \bb. Simultaneously we updated the values of \ttau and \bb used for color correction.
For the $\chi^2$ minimization of the MBB fit we used the same estimates of uncertainties as in Planck XI and Planck XLVIII: the calibration uncertainty (see Table 1 of Planck XI), the associated uncertainty of the CMB removal, and the instrumental noise estimated from the half-difference of the half-mission maps (Planck XLVIII).
We also verified the gain calibration of the detectors by comparing with a scatter plot the detector intensities and MBB model predictions (Figure \ref{fig_calibration}). We included in this plot the COBE/DIRBE map at 60 $\mu$m not used in this work. At this wavelength the emission arising from stochastic heating of very small dust grains causes a departure from a simple MBB form extrapolation \citep{2001ApJ...551..807D}. We note a strong miscalibration of 13\% at 140 $\mu$m. We do not have any explanation for that departure; we obtained a similar departure of 14\% with \ttau, $T$ and \bb taken from Planck  XLVIII.

\subsection{Systematic Uncertainties} \label{systematic_uncertainties}

Along our procedure we introduced various parameters.
We used 8 parameters to define masks latitude and longitude limits and 5 parameters for the mask column density cutoffs. We utilized 2 parameters to define the affine fit limits, 1 for the histogram resolution and 3 corresponding to initial values of $\sigma_{e 353}$, $T$ and $\beta$.
In order to estimate the systematic uncertainties arising from those parameters we repeated 500 realizations of the MBB fit using, instead of their nominal values, a value sampled assuming a normal distribution. In Table \ref{table_systematics} we give the mean and $\sigma$ of those distributions. As guiding principle to estimate the values of $\sigma$, for each parameter we defined a set of possible values based mostly on the observation of residual maps and we attributed the value of $\sigma$ to the standard deviation of this distribution.

As mentioned earlier we also sampled the initial zero-level value for $I_{100\rm{\mu}}$ from a uniform distribution ranging from $\minus$0.1 to 0.2 \Iunits. 

\begin{table*}[]
\hskip-2.0cm\begin{tabular}{cccccc}
                      & Name                    & Mean & $\sigma$ & Units              & Comments                                                                                                                                  \\ \hline
Pegasus-Aquarius mask &                         &       &       &                    &                                                                                                                                           \\
                      & $\sigma_{ini}$          & 9.0   & 0.1   & 10$^{-27}$cm$^{2}$ & Opacity for the $\tau_{353}/\sigma_{e 353}-N_{\text{H}\,\textsc{i}}$ cut                                                                                                                       \\
                      & $C_{DNM \, 2}$          & 1.00  & 0.05  & 10$^{20}$cm$^{-2}$ & Cut in $\tau_{353}/\sigma_{e 353}-N_{\text{H}\,\textsc{i}}$ in the low-resolution map                                                                                                         \\
                      & $b_{min}$               & $\minus$60   & 1     & degree             & Min latitude cut                                                                                                                          \\
                      & $b_{max}$               & $\minus$15   & 2     & degree             & Max latitude cut                                                                                                                          \\
                      & $l_{min}$               & 0     & 3     & degree             & Min longitude cut                                                                                                                         \\
                      & $l_{max}$               & 180   & 3     & degree             & Max longitude cut                                                                                                                         \\
                      & $NH_{cut \, low \, 1}$  & 2.0   & 0.1   & 10$^{20}$cm$^{-2}$ & Low \nhi cut                                                                                                                               \\
                      & $NH_{cut \, high \, 1}$ & 5.0   & 0.25  & 10$^{20}$cm$^{-2}$ & High \nhi cut                                                                                                                              \\
Isothermal mask           &                         &       &       &                    &                                                                                                                                           \\
                      & $NH_{cut \, low \, 2}$  & 2.0   & 0.2   & 10$^{20}$cm$^{-2}$ & Low \nhi cut                                                                                                                               \\
                      & $NH_{cut \, high \, 2}$ & 15.0  & 1.5   & 10$^{20}$cm$^{-2}$ & High \nhi cut                                                                                                                              \\
                      & $\Delta_{eq}$           & 15    & 2     & degree             & Min distance from ecliptic equator                                                                                                        \\
                      & $N_{\sigma}$            & 0.5   & 0.1   & -                  & Half width (in $\sigma$) around $T_{max}$ and $\beta_{max}$                                                                               \\
FUV-NUV Mask          &                         &       &       &                    &                                                                                                                                           \\
                      & $\Delta_{Gould}$        & 30    & 2     & degree             & Half width of Gould Belt region                                                                                                              \\
                      & $b_{cut \, UV}$        & $\minus$40    & 1     & degree             & Latitude cut                                                                                                         \\
Various               &                         &       &       &                    &                                                                                                                                           \\
                      & $\sigma_T$, $\sigma_{\beta}$                        &       & 5     & \%                 & \begin{tabular}[c]{@{}c@{}}Variation of $T$ and $\beta$ for each pixel\\ (for initial color correction and for Isothermal mask)\end{tabular} \\
                      & $N_{bins}$              & 100   & 10    &         -           & Number of bins in histograms                                                                                                              \\
                      & $P_{fit \, min}$       & 5.0   & 0.5   & \%                 & Min percentile for affine fit                                                                                                             \\
                      & $P_{fit \, max}$       & 95    & 2     & \%                 & Max percentile for affine fit                                                                                                             \\
                      & $b_{cut \, zero}$      & 50    & 5     & degree             & Galactic cut for final zero-level monopoles                                                                                              \\
                      & $I_{100\rm{\mu} \, zero}$       &  $\minus$0.1 to 0.2   &   -   & \Iunits             & Initial zero level for $I_{100\rm{\mu}}$                                                                                               \\ 
                     & $NH_{fit \, low \, FUV}$  & 2.5   & 0.5   & 10$^{20}$cm$^{-2}$ & Low \nhi fit limit (FUV)   \\ 
                     & $NH_{fit \, high \, FUV}$  & 6.5   & 0.5   & 10$^{20}$cm$^{-2}$ & High \nhi fit limit (FUV)   \\ 
                     & $NH_{fit \, low \, NUV}$  & 5.   & 0.5   & 10$^{20}$cm$^{-2}$ & Low \nhi fit limit (NUV)   \\ 
                     & $NH_{fit \, high \, NUV}$  & 9.   & 0.5   & 10$^{20}$cm$^{-2}$ & High \nhi fit limit (NUV)   \\ \hline
\end{tabular}
\caption{List of Parameters Sampled in Each of the 500 MBB Fit Realizations.} 
\footnotesize{\textbf{Note.} Parameters sampled from a normal distribution with means and standard deviations given in columns (2) and (3). $I_{100\rm{\mu} \, zero}$ was sampled from a uniform distribution.\\}
\label{table_systematics}
\end{table*}

\subsection{Final Zero Levels}

From the 500 realizations we obtained a distribution of 500 values for the initial and final zero levels, and calibration gains. We fitted those distributions with a Gaussian curve and used the centroids and standard deviations as the final parameters values and uncertainties. The values of the centroids and standard deviations started to converge after only a few tens of realizations.

The final zero levels and uncertainties are shown in the second column of Table \ref{table_zeros}. At 857, 545, and 353 GHz, those final zero levels differ from the initial ones by a couple of percentages. While we randomly sampled between $\minus$0.1 and 0.2 \Iunits the initial zero levels for the 100 $\mu$m map, we found a final zero level of $\minus$0.005$\pm$0.002 \Iunits. This value is compatible with $\minus$0.01$\pm$0.04 \Iunits obtained in the first step from a direct correlation with \nhi. We also provide the calibration gains in the last column of Table \ref{table_zeros}.

\renewcommand{\arraystretch}{1.3}
\begin{table*}[]
\centering                                      
\begin{tabular}{cccccc}
Band    & \multicolumn{3}{c}{Zero Level (MJy sr$^{-1}$)}                                   & Gain  \\ \cline{2-4}
\rule{0pt}{15pt}       & This Work (PR2)  & Planck XLVIII (PR2)  & Planck XI (PR1) &             \\ \hline \hline
353 GHz   &  0.0816 $\pm$ 0.0001  & 0.1248 &   0.085  $\pm$ 0.006 & 0.9998 $\pm$ 0.0001 \\ 
545 GHz   &  0.230 $\pm$ 0.001    & 0.3356   &   0.095 $\pm$ 0.014  & 0.9733 $\pm$ 0.0004  \\ 
857 GHz   &  0.343 $\pm$ 0.002  & 0.5561   &   0.093 $\pm$ 0.009 & 1.0130 $\pm$ 0.0003 \\
240 $\mu$m &  0.55 $\pm$ 0.06  & - &  - & 1.062 $\pm$ 0.002  \\
140 $\mu$m &   0.8 $\pm$ 0.1  & - &  - & 1.126 $\pm$ 0.003  \\
100 $\mu$m &  $\minus$0.005 $\pm$ 0.002  & 0.1128 &   $\minus$0.174 $\pm$ 0.005 & 0.9972 $\pm$ 0.0001  \\
60 $\mu$m  &  0.50 $\pm$ 0.02  & - & - &  2.39 $\pm$ 0.04  \\
\hline
\end{tabular}
\caption{Zero Levels of FIR Maps.}
\footnotesize{\textbf{Note.} The columns are the following, from left to right: frequency or wavelength, final zero levels obtained in this work, zero levels of Planck XLVIII, zero levels of Planck XI, gain calibration resulting from a comparison between observations and MBB predictions (I$_{obs}$=gain$\times$I$_{MBB}$+I$_{zero}$) where I$_{zero}$ is the zero level provided in column (2). Note that because PR2 maps include a CIB monopole one can not directly compare the PR2 zero levels to the PR1 ones.}
\label{table_zeros}
\end{table*}

\subsection{MBB Parameters}\label{MBB_parameters}

\begin{figure*}
  \centering
  \includegraphics[width=0.7\hsize]{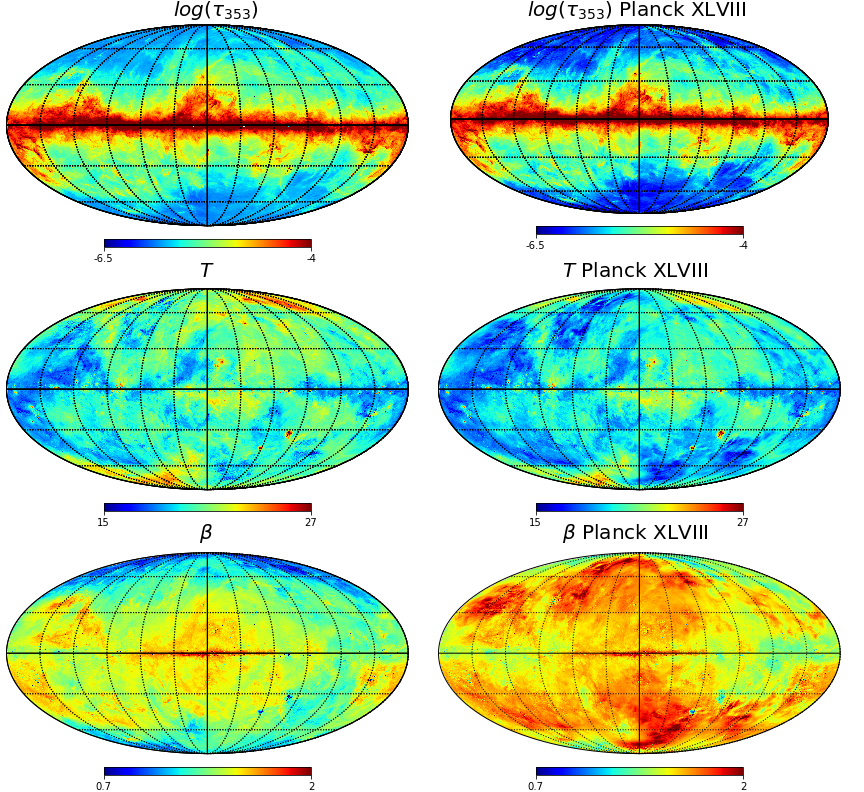}
  \caption{All-sky Mollweide display of the MBB parameters \ttau (top, in log scale), $T$ (middle), and \bb (bottom) extracted from a fit to Planck-HFI and IRAS/DIRBE intensity maps obtained in this work (left) and in Planck XLVIII (right). We superimposed to the maps a 30$^{\circ}$ spaced grid.}
  \label{fig_tau_T_beta}
\end{figure*}

\begin{figure}
  \centering
  \includegraphics[width=\hsize]{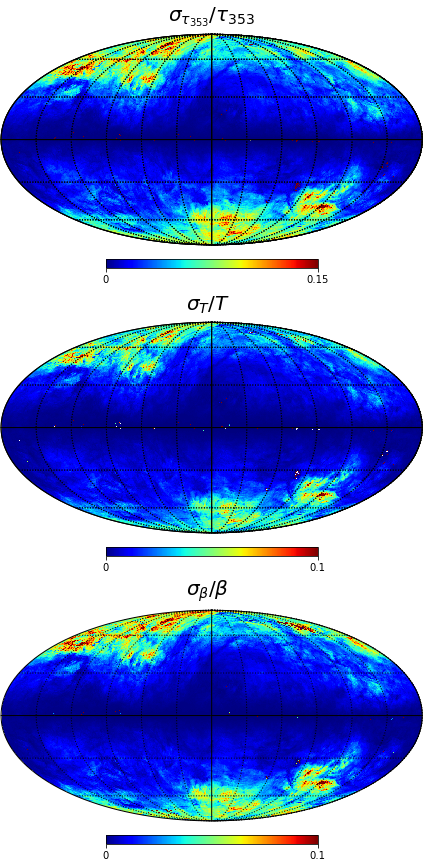}
  \caption{Relative uncertainty of the parameters \ttau (top), $T$ (middle) and \bb (bottom) derived from fits of observations at $100\mu$m and 857, 545, 353 GHz with an MBB spectral form after correction of the zero levels of the intensity maps according to Table \ref{table_zeros}. Each all-sky map displays in Mollweide projection and linear color scale the ratio between the standard deviation of the 500 realizations and its average value. We superimposed to the maps a 30$^{\circ}$ spaced grid.}
  \label{fig_error_params}
\end{figure}

Using the method described above, we obtained for each pixel 500 sets of MBB parameters. In Figure \ref{fig_tau_T_beta} (left), we show the skymap of \ttau, $T$ and \bb averaged over those realizations. For comparison, on the right side of the same figure, we represent the same parameters obtained in Planck XLVIII. In Figure \ref{fig_error_params} we show the associated relative uncertainty based on the standard deviation of the 500 realizations. For the optical depth \ttau, the correction of the zero levels mainly leads to an increase of 7.1$\times$10$^{-7}$ compared to the map of Planck XLVIII. We obtain an average temperature (index) of 20.2 K (1.40)  and an average uncertainty of 0.4 K (0.03). In Planck XLVIII the average temperature (index) was 19.4 K (1.6). The spatial distribution of \bb varied notably from Planck XLVIII, we note a decrease in \bb with Galactic latitude; it also seems that the local increases along the north and south of Loop~I have disappeared. 

\section{Comparison between the Spatial Distributions of $\tau_{353}$ and \nhi}

\subsection{Opacity}

\begin{figure}
  \centering
  \includegraphics[width=\hsize]{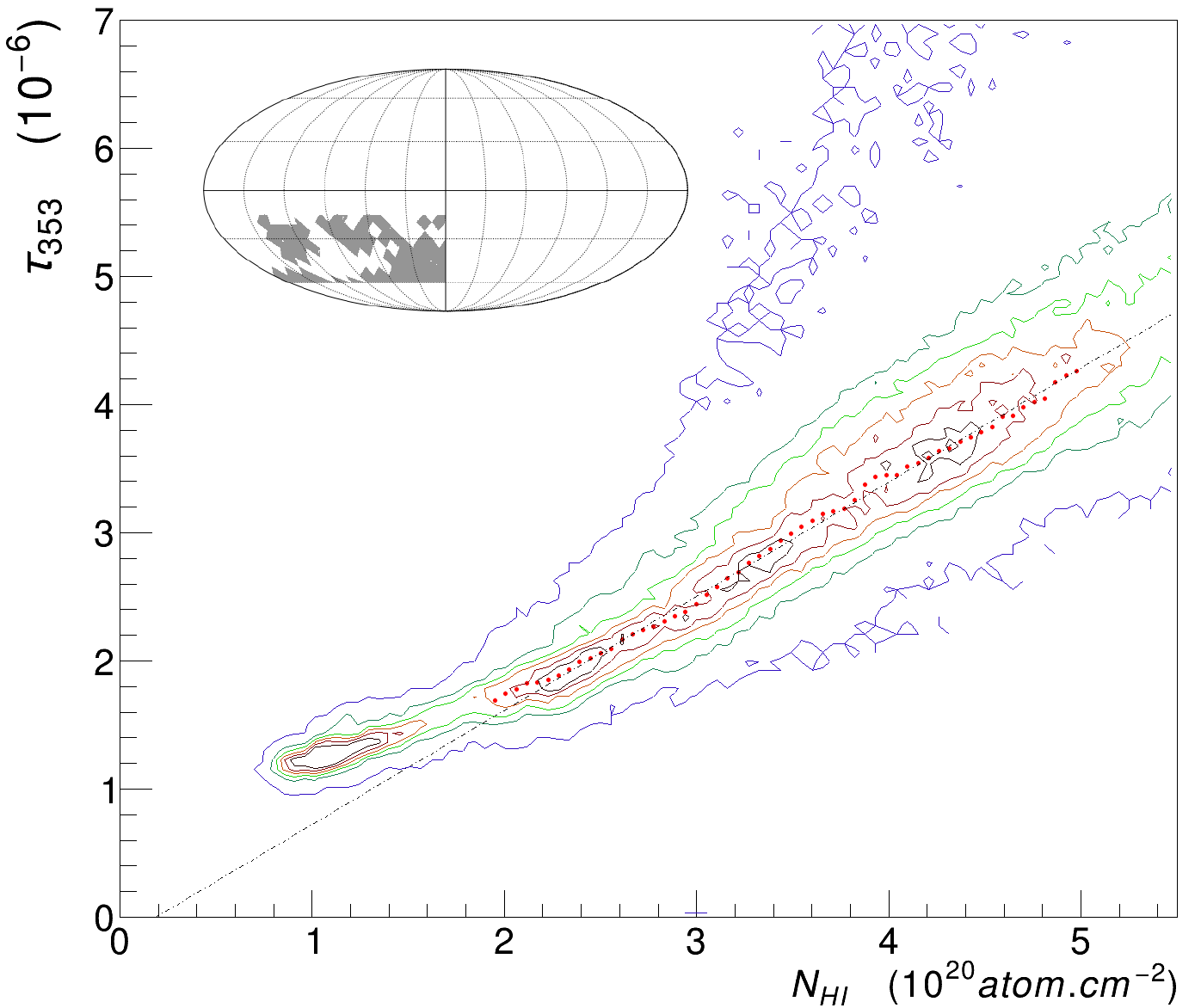}
  \caption{Correlation between \ttau and \nhi. The scatter plot is represented with 6 contours linearly spaced between 1 and 200 pixel counts. The red points are the centroids of the distribution in bins of \nhi for \nhi between 2 and 5$\times 10^{20}$ cm$^{-2}$. We calculated the corresponding error bars, barely visible, by dividing the standard deviation in \ttau by the square root of the number of pixels in each \nhi bin. The dashed line corresponds to the best affine fit of those points. We masked data associated with the white pixels of the Pegasus-Aquarius mask shown in the inset.}
  \label{fig_opacity_scattering}
\end{figure}

We have measured the opacity $\sigma_{e 353}$, the ratio between optical depth and hydrogen column density, in the Pegasus-Aquarius mask with a correlation between \ttau and \nhi (Figure \ref{fig_opacity_scattering}). We observe a linear correlation over a range of \nhi from 2 to 5$\times$10$^{20}$ cm$^{-2}$. As in Figure \ref{fig_zero_857}  we observe for the largest \nhi a tail of overemission of dust compared to hydrogen, which signals a residual presence of gas not traced by optically thin \hi in the mask. In order to remove this bias from the centroid fit, we restricted the upper limit of the \nhi bin in the Gaussian fit to an optical depth corresponding to 1.05 times the centroid position. From an affine fit, we obtain the value $\sigma_{e 353}$=(8.9$\pm$0.1)$\times$10$^{-27}$ cm$^2$ against (6.3$\pm$0.1)$\times$10$^{-27}$ cm$^2$ for Planck XI measured between 1.2 and 2.5$\times$10$^{20}$ cm$^{-2}$. The value of the y-intercept is ($\minus$0.17$\pm$0.07)$\times$10$^{-6}$. 

\begin{figure}
  \centering
  \includegraphics[width=\hsize]{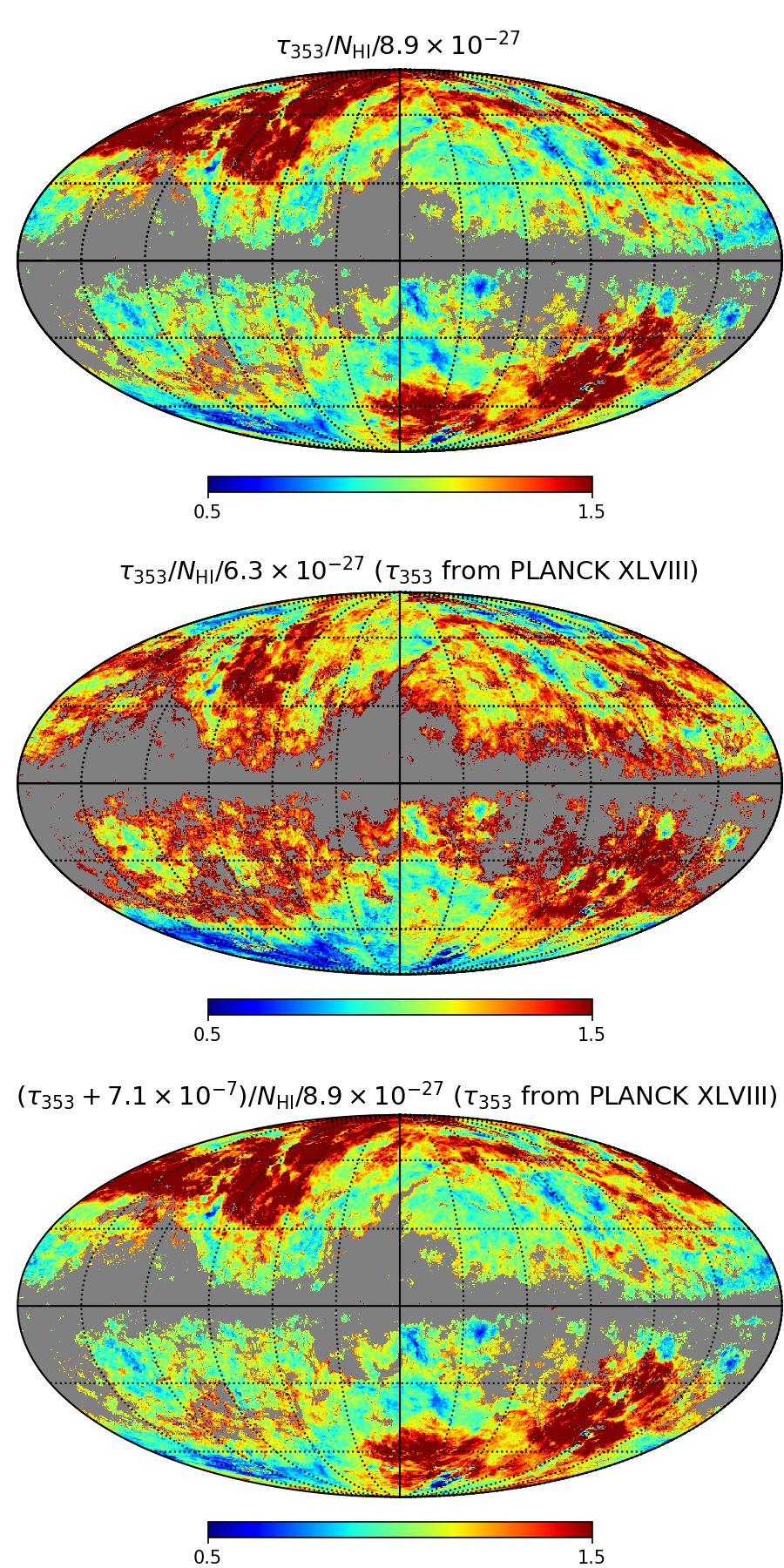}
  \caption{Spatial distribution of the opacity $\sigma_{353 e}=\tau_{353}/N_{\text{H}\,\textsc{i}}$ normalized to the average opacity obtained in this work (top). The middle map shows a similar normalized opacity plot but with \ttau and the average opacity from Planck XLVIII and XI. When adding 7.1$\times$10$^{-7}$ to \ttau derived in Planck XLVIII and dividing by \nhi (bottom), we find a similar opacity map to the one from our work shown at the top of this figure. Maps are displayed in Mollweide projection with a linear color scale and a 30$^{\circ}$ spaced grid. Masked pixels (gray) correspond to regions with $\tau_{353}/\sigma_{e 353}-N_{\text{H}\,\textsc{i}}>2\times10^{20}$ cm$^{-2}$, which removes all regions with potential contamination by H$_2$ or significant \hi self-absorption.}
  \label{fig_opacity_map}
\end{figure}

We show the spatial distribution of the opacity in Figure \ref{fig_opacity_map} (top). We normalized this map to the value of 8.9$\times$10$^{-27}$ cm$^2$ measured in the Pegasus-Aquarius mask.
We masked pixels with $\tau_{353}/\sigma_{e 353}-N_{\text{H}\,\textsc{i}}>2\times10^{20}$ cm$^{-2}$, which removes all regions with potential contamination by H$_2$ or significant \hi self-absorption.
We observe in this map large regions (in green) where the opacity is uniform and equal to the value obtained in the Pegasus-Aquarius mask. More interestingly, we also observe extended regions (in reddish color) of large excesses of dust compared to what would be expected from \nhi assuming the Pegasus-Aquarius mask opacity. Those extended regions correspond to locations where \nhi is the lowest. In Figure \ref{fig_opacity_map} (middle), we plot the same figure but using \ttau from Planck XLVIII normalized to the opacity of 6.3$\times$10$^{-27}$ cm$^2$ measured between 1.2 and 2.5$\times$10$^{20}$ cm$^{-2}$ in Planck XI. As noted by Planck XI, we observe an increase in dust opacity from the diffuse to the denser ISM. This is expected from figure 21 of Planck XI where the smaller opacity obtained with a fit at low \nhi leads to an excess of dust emission at larger \nhi where points systematically depart from the linear fit. It was partially interpreted in Planck XI as a possible increase in dust emissivity. Alternatively an opacity of $\sim$9$\times$10$^{-27}$ cm$^2$ found in this paper nicely fits the data points above 2$\times$10$^{20}$ cm$^{-2}$ in Figure 21 of Planck XI but leads to large excesses of dust at low \nhi. The increase in opacity observed in Planck XI at intermediate \nhi may suffer from incorrect zero levels that lead to an offset in \ttau. If we add to \ttau derived in Planck XLVIII the constant 7.1$\times$10$^{-7}$ (Figure \ref{fig_opacity_map}, bottom), we find the same opacity spatial distribution as in our work (Figure \ref{fig_opacity_map}, top). While the opacity is mostly equal to or larger than the average opacity, we note in Figure \ref{fig_opacity_map} (top) that for longitude between 60$^{\circ}$ and 180$^{\circ}$ and latitude less than $\minus$60$^{\circ}$ we observe a large region of lower opacity. This is likely due to an oversubtraction of the zodiacal emission in the dust maps.

\begin{figure}
  \centering
  \includegraphics[width=\hsize]{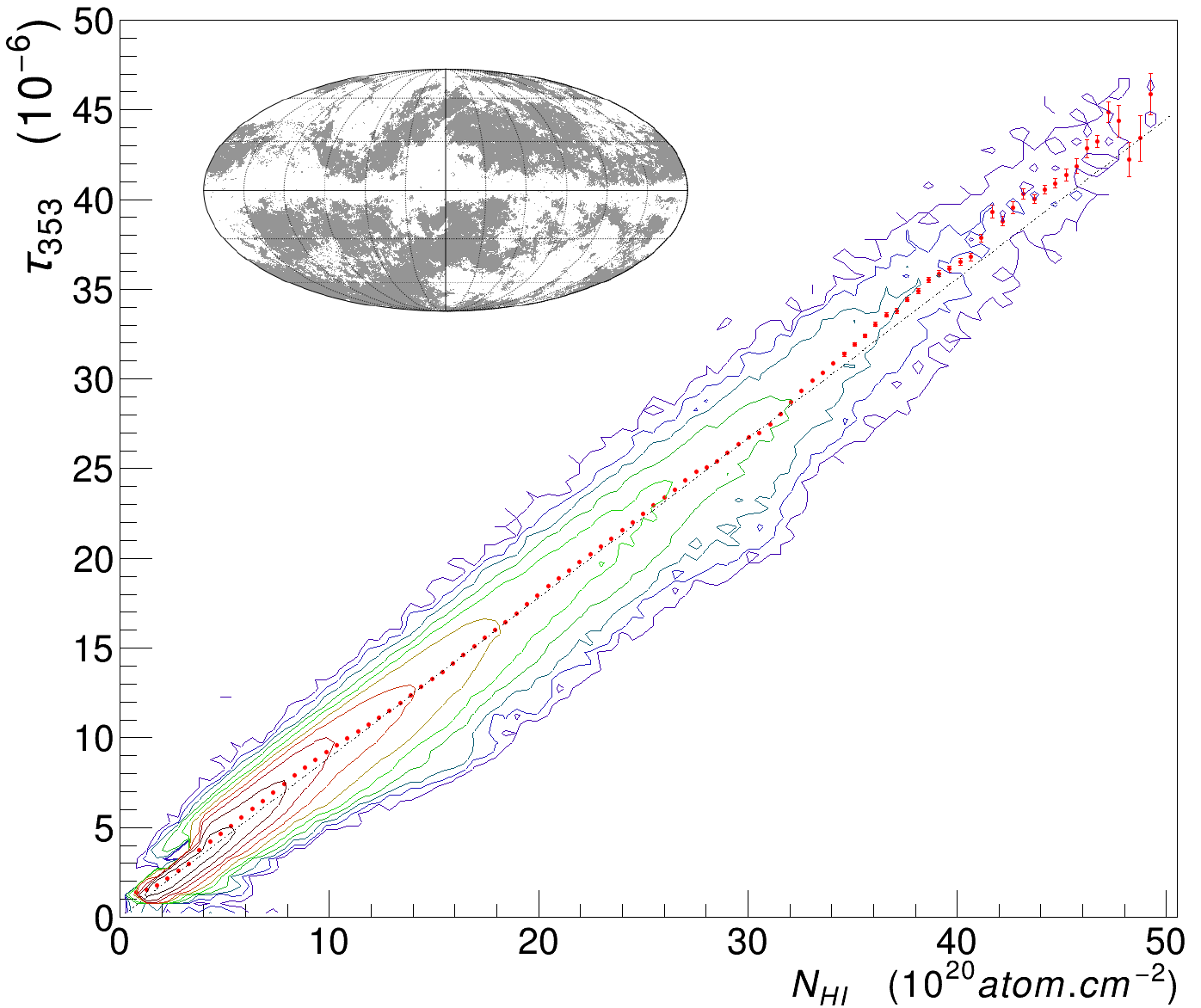}
  \caption{Correlation plot between \ttau and \nhi displayed with 10 contours spaced in log between 2 and 25,000 pixel counts. The red points correspond to the centroids of the \nhi bins. We calculated the corresponding error bars, barely visible, by dividing the standard deviation in \ttau by the square root of the number of pixels in each \nhi bin. The dashed line corresponds to $\sigma_{e 353}$=8.9$\times$10$^{-27}$ cm$^2$, average opacity extracted from the smaller Pegasus-Aquarius mask. We masked data associated with the white pixels of the mask shown in the inset and described in the text.}  
  \label{fig_opacity_large}
\end{figure}

Similarly to what we did in Figure \ref{fig_opacity_scattering} in the small Pegasus-Aquarius mask, we expect a linear correlation between \ttau and \nhi in a wider section of the sky if we exclude pixels 
with potential contamination by H$_2$ or significant \hi self-absorption, and pixels with very low \nhi where we observed the dust excesses of Figure \ref{fig_opacity_map} (top). Those regions of dust excess can be characterized by an excess of optical depth  $\tau_{353}-N_{\text{H}\,\textsc{i}}\times\sigma_{e 353}$ larger than 0.03$\times$10$^{-6}$ and \nhi less than 4$\times$10$^{20}$ cm$^{-2}$. If we exclude those pixels and pixels with $\tau_{353}/\sigma_{e 353}-N_{\text{H}\,\textsc{i}}>2\times10^{20}$ cm$^{-2}$ (inset of Figure \ref{fig_opacity_large}), we observe an excellent linear proportionality between \ttau and \nhi  over a wide range of \nhi (Figure \ref{fig_opacity_large}). We represented the average opacity of $\sigma_{e 353}$=(8.9$\pm$0.1)$\times$10$^{-27}$ cm$^2$  measured in the Pegasus-Aquarius mask by the dotted line in Figure \ref{fig_opacity_large}. We note a departure from this value for \hi column densities above $\sim$35$\times$10$^{20}$ cm$^{-2}$ probably caused by the presence of DNM in the mask at high column density or when the \hi that starts to be opaque (not corrected here). 

We can also express \ttau in terms of absorption cross-section per dust mass $\kappa_{353}$ at 353 GHz: $\tau_{353}=M_{dust}\kappa_{353}$ where the dust mass column density $M_{dust}$ depends on the dust-to-gas mass ratio $D/G$ and on the gas mass per H-atom $\mu$=1.4 ($M_{dust}=\mu m_{\rm{H}} N_{\rm{H}} D/G$ where $m_{\rm{H}}$ is the mass of a hydrogen atom). We can then deduce the dust-to-gas mass ratio: $D/G=\sigma_{e 353}/(\mu m_{\rm{H}}\kappa_{353})$. Using for $\kappa$ the power-law approximation of the dust model of \cite{2017A&A...602A..46J} obtained by \cite{2018ARA&A..56..673G}, we have $\kappa=0.64(\lambda/250)^{-1.79}$ cm$^2$g$^{-1}$ that gives $D/G$=0.53$\times$10$^{-2}$ with $\sigma_{e 353}$=(8.9$\pm$0.1)$\times$10$^{-27}$cm$^2$. Note that the main uncertainty in this number resides in the value of $\kappa_{353}$ (see Figure 4a of \cite{2018ARA&A..56..673G} ) and not in the value of the average opacity.

\subsection{Spatial Distribution of the Dust Excesses}

\begin{figure*}
  \centering
  \includegraphics[width=\hsize]{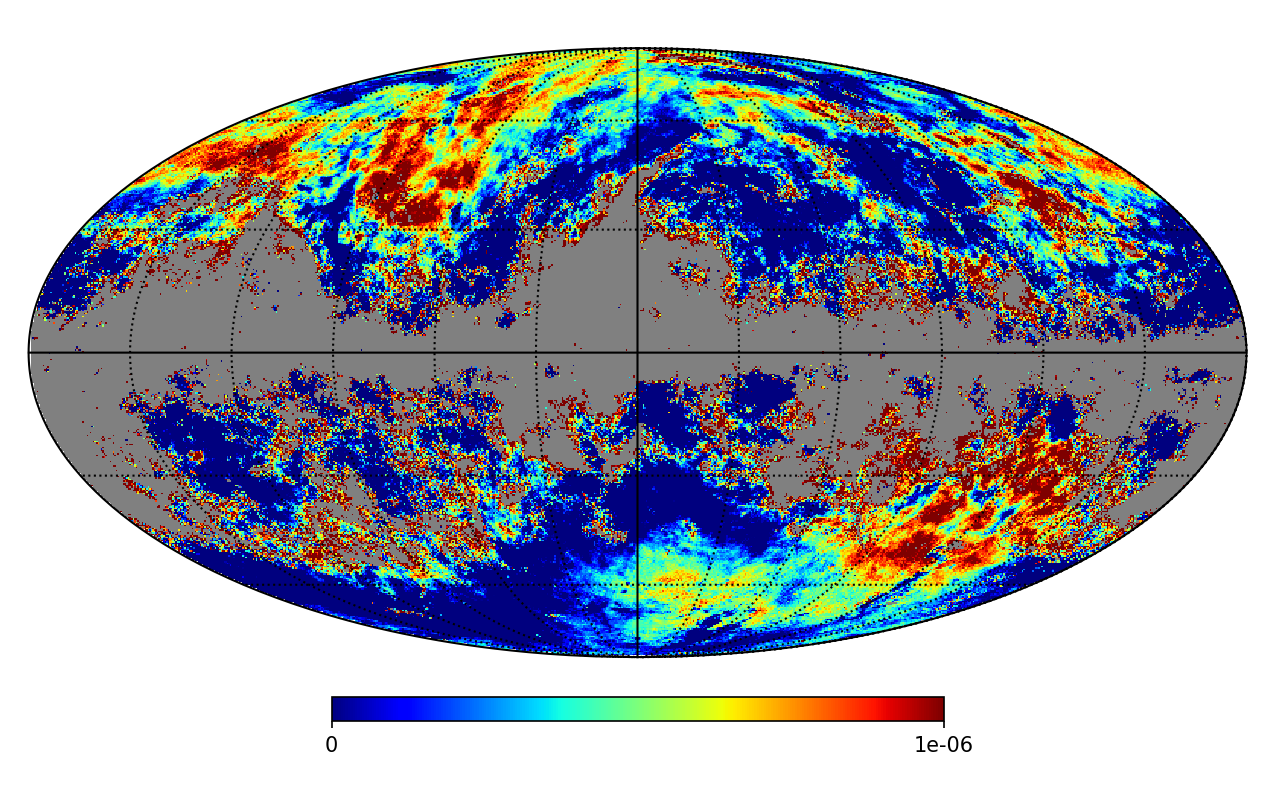}
  \caption{All-sky Mollweide display of the excess of optical depth at 353 GHz compared to \nhi predictions: $\tau_{353}-N_{\text{H}\,\textsc{i}}\times\sigma_{e 353}$. We used the opacity $\sigma_{e 353}$=8.9$\times$10$^{-27}$ cm$^2$ measured in the Pegasus-Aquarius mask. The color is scaled linearly with the map intensity, and a 30$^{\circ}$ spaced grid is superimposed. We masked pixels with $\tau_{353}/\sigma_{e 353}-N_{\text{H}\,\textsc{i}}>2\times10^{20}$ cm$^{-2}$  where $\tau_{353}$ is not a good tracer for \nhi.}
  \label{fig_dcm}
\end{figure*}

We derived the residual map of the optical depth by subtracting from \ttau the product of the opacity, derived in the Pegasus-Aquarius mask, with \nhi. We show the skymap of the optical depth residuals in Figure \ref{fig_dcm}. Outside the regions with potential contamination by H$_2$ or significant \hi self-absorption, in which correction for \hi optical thickness and CO proxy prevent an accurate census of the gas, the residuals should be fluctuating around 0. As we noted for the opacity map, we observe a large excess of dust emission above the emission associated to hydrogen traced by \hi. The optical depth of this extra dust is $\sim$0.5-1$\times$10$^{-6}$. It is mainly located in the northeast part of the skymap ($l$=60$^{\circ}$-180$^{\circ}$ and $b$=30$^{\circ}$-90$^{\circ}$) and in the southwest ($l$=180$^{\circ}$-360$^{\circ}$ and $b$=0$^{\circ}$-90$^{\circ}$). 

To derive \nhi from the brightness temperature, we used an optically thin approximation. The dust excess cannot result from this approximation since, at low \nhi, a spin temperature of for example 150K would lead to an \nhi increase of only a few percent. In comparison we observe an excess of up to a factor 2.5 toward the Lockman Hole.
Note that this excess was not observed previously (\cite{1998ApJ...500..525S, 2014A&A...566A..55P}; Planck XI; Planck XLVIII) because it was implicitly erased by calculating the zero (detector offset and/or extragalactic emission) of the dust map from a correlation with \nhi at very low \nhi.

\begin{figure}
  \centering
  \includegraphics[width=\hsize]{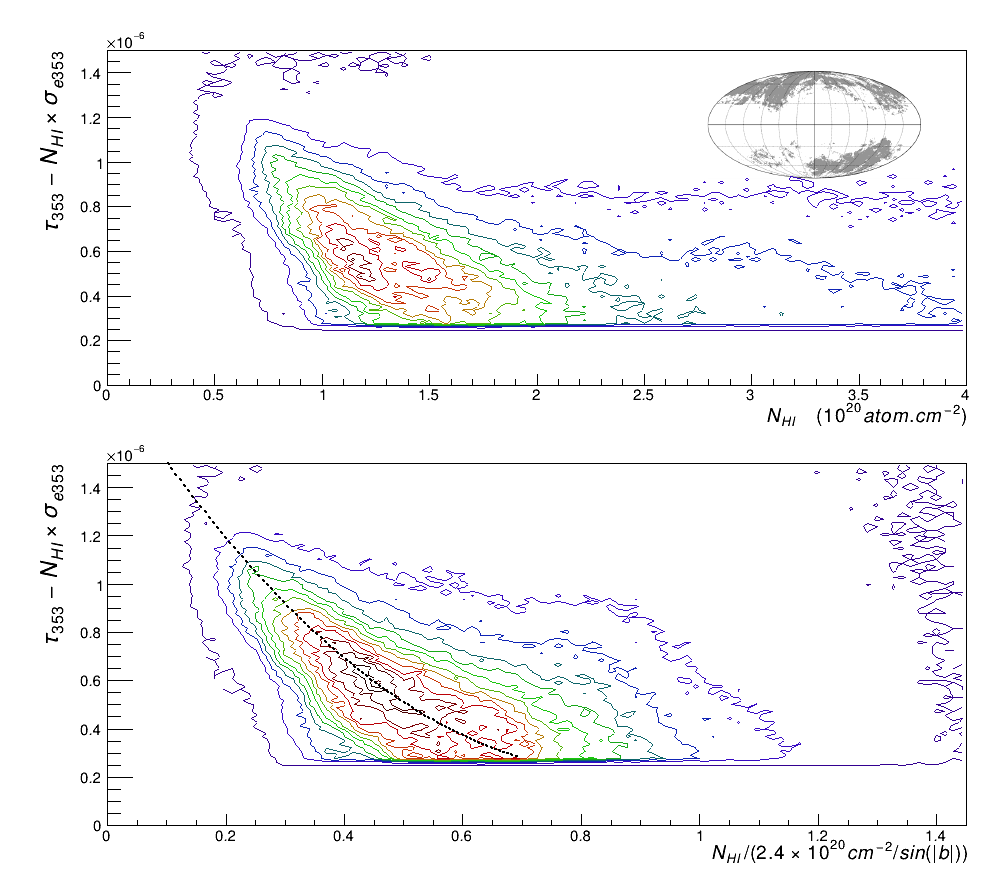}
  \caption{
     Correlation of the optical depth of the dust excess and \nhi (top) and the optical depth of the dust excess and $N_{\text{H}\,\textsc{i}}$/(2.4$\times$10$^{20}$ cm$^{-2}$/sin($|$b$|$)) (bottom). The dashed line corresponds to the function (2.5X$^2$ - 4.3X + 2.1)$\times$10$^{20} \times \sigma_{e 353}$ where X=$N_{\text{H}\,\textsc{i}}$/(2.4$\times$10$^{20}$ cm$^{-2}$/sin($|$b$|$)), which represents a good approximation for the optical depth of the excess. The scatter plots have 13 contours linearly spaced between 1 and 700 pixel counts. We masked data associated to white pixels of mask shown in the inset. Gray pixels correspond to a dust excess larger than 0.03$\times$10$^{-6}$ and \nhi less than 4$\times$10$^{20}$ cm$^2$.}
  \label{fig_anticorrel}
\end{figure}

In Figure \ref{fig_anticorrel} (top), we show the anticorrelation between the excess of dust and \nhi. This anticorrelation naturally results from the monopole of 7.1$\times$10$^{-7}$ in \ttau (see Section \ref{MBB_parameters}), itself related to the new values of the zero levels described earlier. We obtain a stronger anticorrelation if we use, instead of \nhi the variable $N_{\text{H}\,\textsc{i}}$/(2.4$\times$10$^{20}$ cm$^{-2}$/sin($|$b$|$)) (Figure \ref{fig_anticorrel}, bottom), which highlights the \nhi contrast in the local ISM over the 1/sin($|$b$|$) trend expected for longer lines of sight crossing the \hi disk.
We extracted the normalization coefficient 2.4$\times$10$^{20}$ cm$^{-2}$ from a fit to \nhi.
The stronger anticorrelation is materialized by higher pixel counts in the central part of the scatter plot of Figure \ref{fig_anticorrel} (bottom) as well as a narrower distribution.

\cite{2006ApJ...637..333H} measured toward the globular cluster Messier 3 (l=42.2$^{\circ}$, b=78.7$^{\circ}$) a strong depletion of iron compared to sulfur in both the warm neutral medium (WNM) and WIM. Since iron is heavily incorporated into dust grains in contrast to sulfur, one expects the presence of dust associated to the WIM in that direction. This is what we may observe since this line of sight intercepts the dust excess region.

\begin{figure}
  \centering
  \includegraphics[width=\hsize]{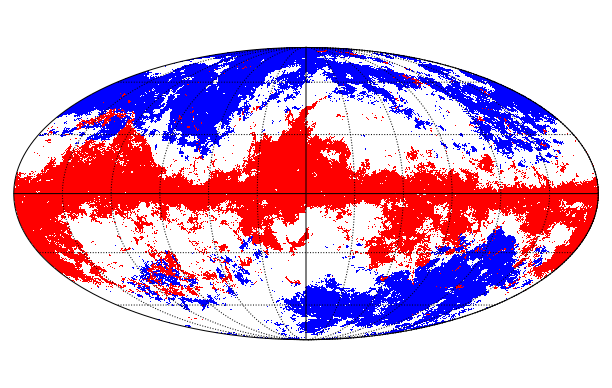}
  \caption{{All-sky map displaying the partition of the sky in 3 large-scale regions. The dust excess region is represented with blue pixels (25\% of the sky) while red ones correspond to $\tau_{353}/\sigma_{e 353}-N_{\text{H}\,\textsc{i}}>2\times10^{20}$ cm$^{-2}$  where $\tau_{353}$ is not a good tracer for \nhi (30\% of the sky). In the region with white pixels FIR dust emission and \hi 21 cm emission linearly correlate (45\% of the sky). The map is centered on the Galactic Center and is displayed in Mollweide projection with a 30$^{\circ}$ spaced grid.}}
  \label{fig_3regions}
\end{figure}
 
Dust or gas studies outside the plane are often based on simple latitude cuts. In this paper, we define instead three regions based on the comparison between \hi, CO, and FIR emission: a region of DNM and high column density gas (red pixels in Figure \ref{fig_3regions}) where the optical depth of the dust deduced from FIR emission is large and larger than those expected from \nhi (traced by its 21 cm line) and $N_{\rm{H}_2}$ (traced by the CO molecule), a region (that we called in this work the excess region corresponding to blue pixels in Figure \ref{fig_3regions}) where the optical depth of the dust is small but larger than that expected from \nhi, and finally a region (white pixels in Figure \ref{fig_3regions}) where FIR dust emission and \hi 21 cm emission linearly correlate.

\section{Dust and Reynolds Layer of the WIM}

\cite{1999A&A...344..322L} observed an excess of dust associated to H$^+$ from the WIM through a cosecant latitude dependence of the residual emission of dust detected by DIRBE/FIRAS and COBE/DIRBE once dust associated to \hi was subtracted. It was also observed in \cite{2000A&A...354..247L} from a decomposition of dust emission into a component associated with \nhi and another with H$_{\alpha}$ emission. This observation was not confirmed by \cite{2007ApJ...667...11O}.

  We searched for dust associated to the Reynolds layer of the WIM. Similarly to what we did in the Pegasus-Aquarius mask we fitted \ttau derived in this work with a linear combination of \nhi, isotropic, and \nhplus templates predicted by NE2001 and from H$_{\alpha}$ intensity. We calibrated the H$_{\alpha}$ emission using \nhplus from NE2001. We found a good correlation with $I_{NE2001}$=($I_{H_{\alpha}}$/1.3R + 0.6)$\times$10$^{20}$ cm$^{-2}$ up to an H$_{\alpha}$ intensity of 3 Rayleigh. Above this value, the H$_{\alpha}$ emission is dominated by bright \hii regions. We performed the linear combination fit in the same mask as in Figure \ref{fig_opacity_large} additionally masking pixels of absolute ecliptic latitudes smaller than 15$^{\circ}$ or with an H$_{\alpha}$ intensity larger than 3 R. The fit set the opacity associated with the NE2001 template to a value close to 0, corresponding to a dust abundance 130 times smaller in H$^+$ than in HI. With the H$_{\alpha}$ template, we found a larger detection corresponding to a dust abundance 14 times smaller in H$^+$ than in HI. However this marginal detection might be due to some residual contamination of free-free emission in the Planck dust maps.
Our unsuccessful detection of dust associated to H$^+$ in the Reynolds layer of the WIM does not necessarily imply that this component is dust free, but rather that its FIR emission is below our detection threshold because the \nhplus column densities are small compared to the \nhi ones, and because they partially correlate with the large HI column densities in the widespread WNM.

\section{Far- and Near-ultraviolet Diffuse Emission}

At FUV and NUV wavelengths, the diffuse light is dominated by the scattering of UV-bright stars on interstellar dust grains \citep{1979ApJ...227..798J}; at high latitudes, FUV and NUV intensities are then expected to scale with dust. \cite{2016MNRAS.459.1710M} calculated that $\sim$25\% of the scattered light arises from 10 bright stars, and half of the diffuse radiation originates within $\sim$200 pc. Additionally, and particularly along the Gould Belt where most UV-bright stars reside, the FUV and NUV emissions include stellar sources \citep{2015ApJ...798...14H}. Figures \ref{fig_opacity_map} (top) and \ref{fig_dcm} suggest the existence of large dust excesses compared to the quantity expected from \nhi. This excess of dust should also scatter UV photons and should lead to FUV and NUV excesses compared to predictions based on \nhi.

\subsection{UV Data}

We analyzed FUV and NUV diffuse maps\footnote{hlsp\_uv-bkgd\_galex\_map\_allsky\_fuv\_v1\_skymap.fits and hlsp\_uv-bkgd\_galex\_map\_allsky\_nuv\_v1\_skymap.fits} extracted by \cite{2014ApJS..213...32M} from the General Releases 6 and 7 of the Galaxy Evolution Explorer (GALEX) spacecraft \citep{2005ApJ...619L...1M} at 1350–1750 $\AA$ (FUV) and 1750-2850 $\AA$ (NUV). We transformed those point-source free maps downloaded from the Mikulski Archive for Space Telescopes\footnote{\url{https://archive.stsci.edu/prepds/uv-bkgd/}} with a resolution of 2' into HEALPix standard with a resolution parameter $N_{side}=512$. At this resolution, the pixelization of GALEX is washed-out, and both FUV and NUV maps have a smooth spatial distribution. Both FUV and NUV maps  were corrected for airglow from geocoronal oxygen. The zodiacal light was subtracted from the NUV map. This contribution is negligable in the FUV skymap because it arises from sunlight scattered off dust in the ecliptic plane and because the solar flux is strongly suppressed at FUV wavelength. 

To analyze those UV skymaps, we masked, in addition to high intensity regions not observed by GALEX, pixels within 40$^{\circ}$ of the Gould Belt using the parameterization of \cite{2019Ap&SS.364...19C}, or with latitude comprised between $\minus$40$^{\circ}$ and 0$^{\circ}$, or 10$^{\circ}$ from the bright stars Spica, Achernar, Algenib, Sirius, Mirzam, and Adhara known to possess dust-scattered UV halos \citep{2011ApJ...734...13M}. Additionnally we selected bright UV stars from the Thor-Delta (TD1) catalog \citep{ 1995yCat.2059....0T} with a flux at 1565 $\AA$ above 10$^{-10}$ erg cm$^{-2}$s$^{-1}$$\AA$$^{-1}$ and we masked pixels within 2$^{\circ}$.5 from sources position. We also masked pixels where \nhi is not a good proxy for the dust column density:
we first excluded pixels with $\tau_{353}/\sigma_{e 353}-N_{\text{H}\,\textsc{i}}>2\times10^{20}$ cm$^{-2}$;
we then masked pixels corresponding to the excess of dust shown in Figure \ref{fig_dcm} and characterized by an excess of optical depth  $\tau_{353}-N_{\text{H}\,\textsc{i}}\times\sigma_{e 353}$ larger than 0.03$\times$10$^{-6}$ and \nhi less than 4$\times$10$^{20}$ cm$^{-2}$. We show the resulting mask in the insets of Figure \ref{fig_correlation_UV_NHI}a and \ref{fig_correlation_UV_NHI}b.

\subsection{Comparison with $N_{\text{H}\,\textsc{i}}$}

\begin{figure}
  \gridline{
    \fig{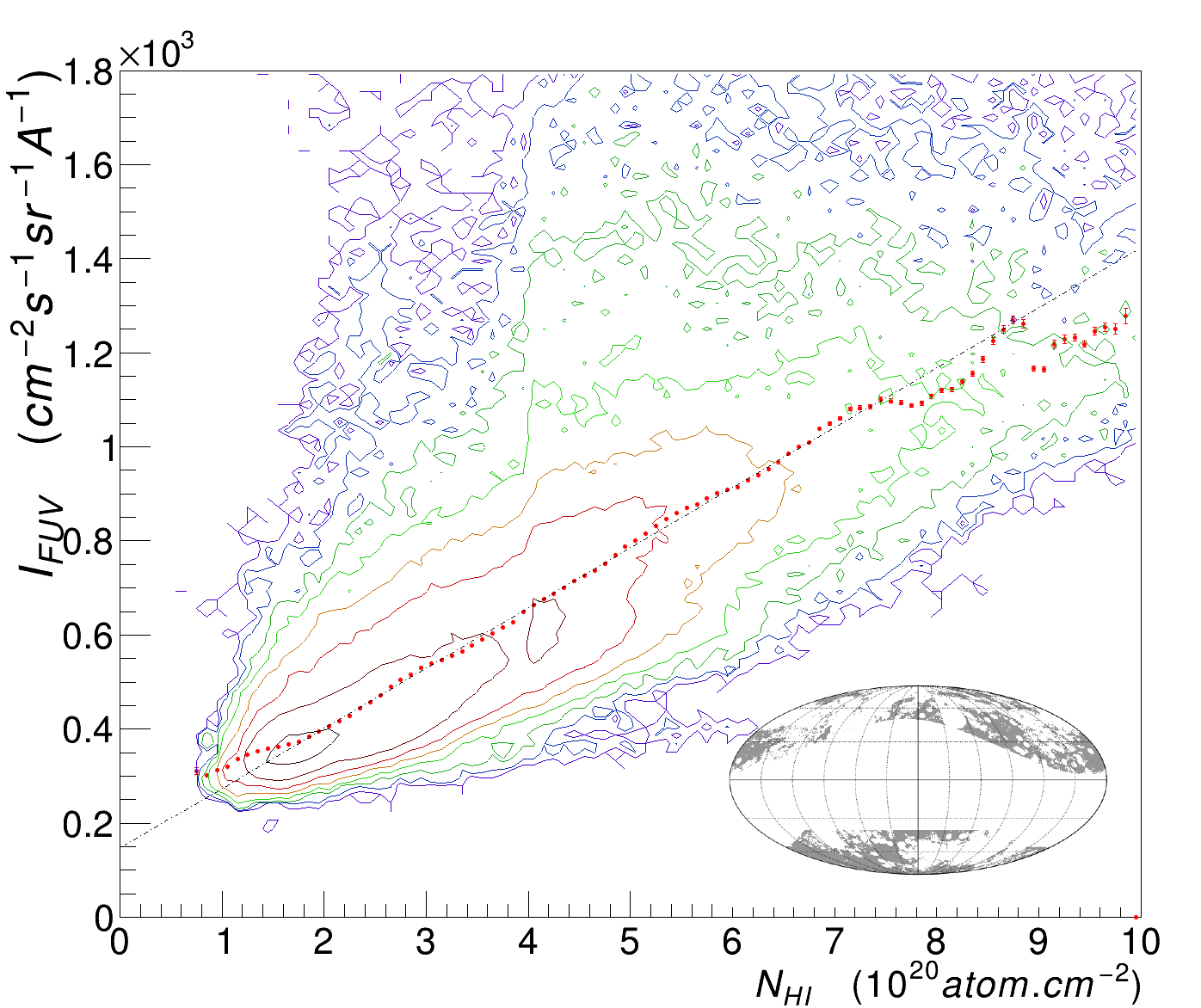}{.25\textwidth}{a}
    \fig{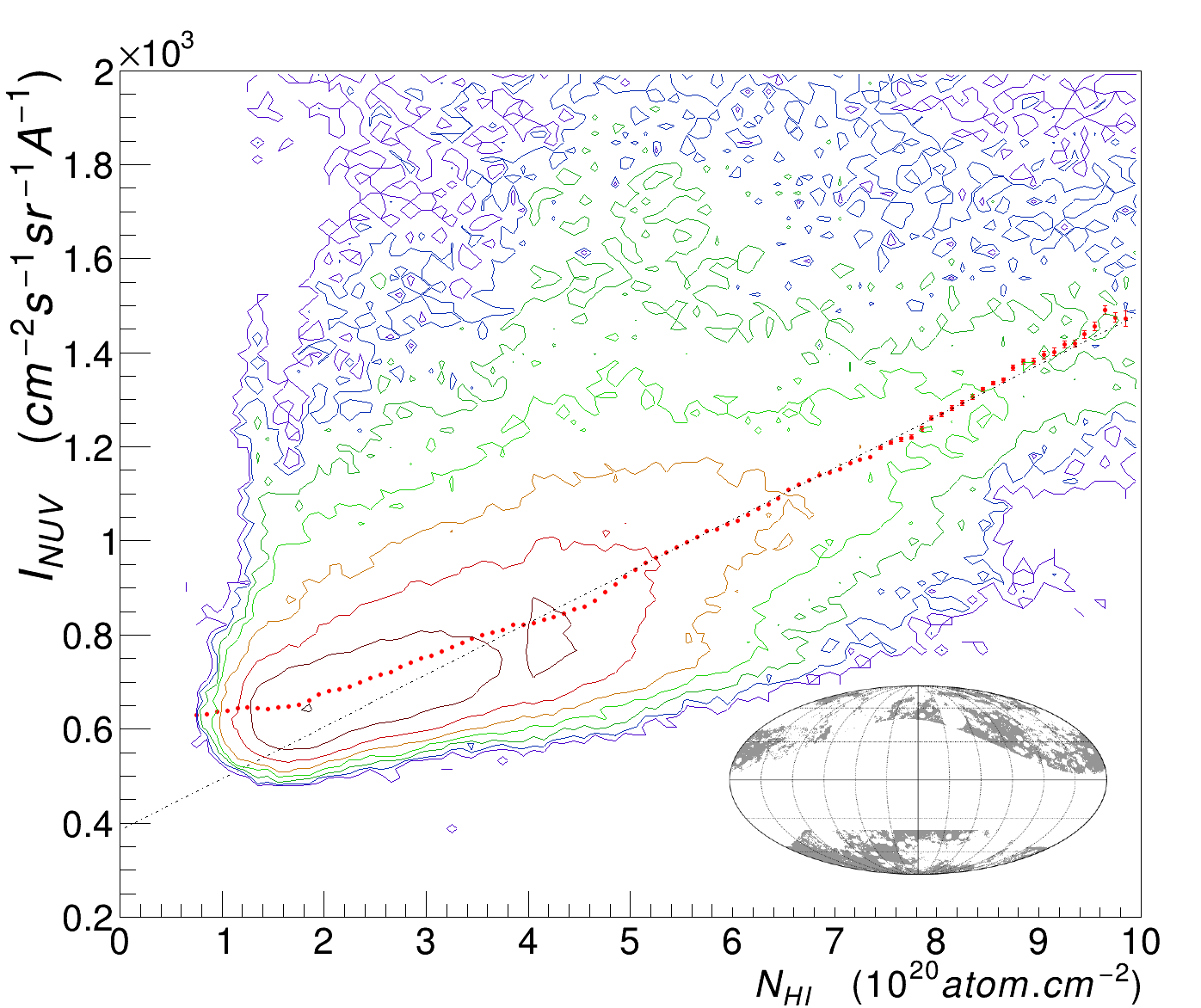}{.25\textwidth}{b}}
\gridline{\fig{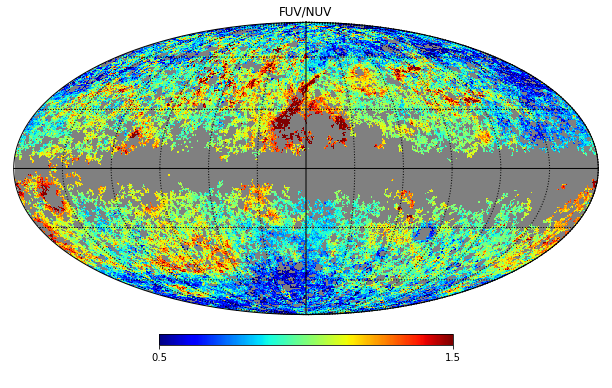}{.4\textwidth}{c}}
\caption{Correlation between local \nhi and GALEX FUV intensity $I_{FUV}$ (a) and NUV intensity $I_{NUV}$ (b) displayed with 8 contours linearly spaced between 1 and 1500 pixel counts. The red points are the centroids of the distribution in bins of \nhi. We calculated the corresponding error bars, barely visible, by dividing the standard deviation in UV intensity by the square root of the number of pixels in each \nhi bin. The dashed line corresponds to the best affine fit to the points calculated in the range from 2.5 to 6.5$\times$10$^{20}$ cm$^{-2}$ for plot (a) and 5 to 9$\times$10$^{20}$ cm$^{-2}$ for plot (b). We masked data associated with the white pixels of the mask shown in the inset and detailed in the text. The ratio between $(I_{FUV}-I_{0(FUV)})$ and $(I_{NUV}-I_{0(NUV)})$ divided by their respective opacities is displayed in (c).
\label{fig_correlation_UV_NHI}}
\end{figure}

Similarly to what we did at FIR wavelengths, we compared the GALEX FUV and NUV diffuse intensities with \nhi inside the mask described above. With UV light being absorbed by the dust, GALEX did not detect radiation emitted at distances farther than $\sim$600 pc \citep{2016MNRAS.459.1710M} so here we considered only \hi with a radial Galactic distance comprised between 7 and 9 kpc. Figure \ref{fig_correlation_UV_NHI} shows the scatter plot of I$_{FUV}$ and I$_{NUV}$, the GALEX FUV and NUV intensities, and local \nhi. We observe at both FUV and NUV wavelengths an affine correlation between the two observables with an average Galactic FUV intensity per \nhi (that we call average FUV opacity by analogy with the FIR terminology) of (130$\pm$4)$\times$10$^{-20}$ photons s$^{-1}$cm$^{-2}$sr$^{-1}$$\AA$$^{-1}$cm$^{2}$ and an average Galactic NUV intensity per \nhi (average NUV opacity) of (111$\pm$6)$\times$10$^{-20}$ photons s$^{-1}$cm$^{-2}$sr$^{-1}$$\AA$$^{-1}$cm$^{2}$. The FUV opacity agrees with the value of (121$\pm$30) $\times$10$^{-20}$ photons s$^{-1}$cm$^{-2}$sr$^{-1}$$\AA$$^{-1}$cm$^{2}$ found by \cite{2013ApJ...779..180H} for latitudes with absolute values between 30$^{\circ}$ and 45$^{\circ}$, but it differs from the smaller FUV opacities they found at higher absolute value latitudes.
The y-intercept of the FUV scatter plot is $I_{0(FUV)}$=137$\pm$15 photons s$^{-1}$cm$^{-2}$sr$^{-1}$$\AA$$^{-1}$ and the one of the NUV plot is I$_{0(NUV)}$=378$\pm$45 photons s$^{-1}$cm$^{-2}$sr$^{-1}$$\AA$$^{-1}$.
We observe in the NUV scatter plot of Figure \ref{fig_correlation_UV_NHI}b a strong departure from linearity for \nhi less than 4$\times$10$^{20}$ cm$^{-2}$. This departure may be due to zodiacal light not totally removed from the NUV map (see next paragraph). We calculated the uncertainties for those values using the same method as in the FIR sections; we derived 500 FUV scatter plots and 500 NUV scatter plots with randomly sampled mask and fit limits given in Table \ref{table_systematics}.

\begin{figure}
  \centering
  \includegraphics[width=\hsize]{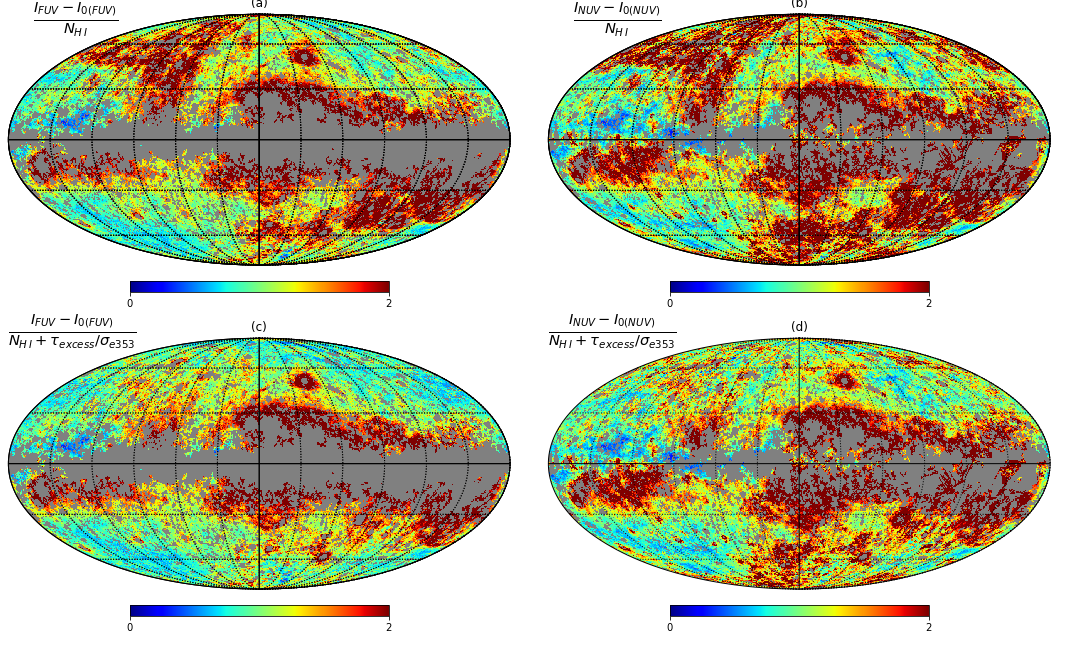}
  \caption{ Spatial distribution of (a) the FUV opacity ($I_{FUV}-I_{0(FUV)})/N_{\text{H}\,\textsc{i}}$ /130$\times$10$^{-20}$ and (b) the NUV opacity $(I_{NUV}-I_{0(NUV)})/N_{\text{H}\,\textsc{i}}$ /111$\times$10$^{-20}$ where we subtracted from UV intensities the offsets corresponding to the y-intercept of the dashed line displayed in Figure \ref{fig_correlation_UV_NHI} and normalized the ratio by its slope, average UV intensity per \nhi. In the lower row, plots (c) and (d) represent the same quantities as (a) and (b), but we added to \nhi  the excess of optical depth observed at FIR frequencies divided by $\sigma_{e 353}$. We defined this dust excess as $\tau_{353}-N_{\text{H}\,\textsc{i}}\times\sigma_{e 353}$ larger than 0.03$\times$10$^{-6}$ and \nhi less than 4$\times$10$^{20}$ cm$^{-2}$. In the labels of the plots, we named this optical depth excess $\tau_{excess}$  and we omitted the normalization factors. Maps are displayed in Mollweide projection with a linear color scale and a 30$^{\circ}$ spaced grid. Masked pixels (gray) correspond to regions not observed by GALEX. }
  \label{fig_FUV_opacity_map}
\end{figure}

We present in Figure \ref{fig_FUV_opacity_map} (upper row) the spatial distribution of the FUV opacity $(I_{FUV}-I_{0(FUV)})$/$N_{\text{H}\,\textsc{i}}$ and of the NUV opacity $(I_{NUV}-I_{0(NUV)})$/$N_{\text{H}\,\textsc{i}}$ normalized to the average values obtained by the fit of Figure \ref{fig_correlation_UV_NHI}. We observe large-scale excesses of FUV and NUV emission compared to what we expect from \nhi.
One does not expect a good correlation between the diffuse UV photons and \nhi close to the Galactic plane and along the Gould Belt or close to bright UV stars. But we also note excesses in the northeast and southwest part of the maps with distributions similar to those obtained in Figure \ref{fig_opacity_map} (top) and in Figure \ref{fig_dcm}.
This seems to indicate that the excess observed in the FIR domain corresponds to an excess of dust column density rather than an increase of thermal dust emissivity at low \nhi.
We do not observe major FUV emission from dust associated to molecular clouds or to regions of high DNM shielded from UV. In Figure \ref{fig_FUV_opacity_map} (lower row), we show the same FUV and NUV opacities, but we added to the local \nhi the excess of optical depth observed at FIR frequencies divided by $\sigma_{e 353}$. As we did previously, we defined this dust excess as $\tau_{353}-N_{\text{H}\,\textsc{i}}\times\sigma_{e 353}$ larger than 0.03$\times$10$^{-6}$ and \nhi less than 4$\times$10$^{20}$ cm$^{-2}$. When this contribution is added to the local \nhi, the northeast and southwest FUV and NUV excesses vanish. We conclude that those northeast and southwest excesses (compared to \nhi) observed in FIR, FUV, and NUV are compatible both spatially and in intensity.
We observe a residual contribution in the NUV map of Figure \ref{fig_FUV_opacity_map}d for central longitudes and latitude lower than $\minus$45$^{\circ}$. This contribution is absent from the FUV skymap. It is the main origin of departure from linearity observed at low \nhi in Figure \ref{fig_correlation_UV_NHI}(b).
It could be due to residual zodiacal light at NUV wavelength, but its limited extent in ecliptic longitude is puzzling.

\cite{2019MNRAS.489.1120A} (Figure 3) observed an excess of FUV emission for pixels with latitudes between 70$^{\circ}$ and 80$^{\circ}$ and with low dust FIR intensities. Since this region intercepts our dust excess, the excess they observed has probably the same origin as ours. Based on the fact that they did not observe a similar excess at NUV wavelength, they proposed that it could originate from Lyman-band fluorescent emission from molecular hydrogen instead of dust scattering. We do not confirm this hypothesis because we also observed this excess at NUV wavelength. Additionally we do not see any large-scale excess similar to that of our FIR excess (Figure \ref{fig_opacity_map} (top) and in Figure \ref{fig_dcm}) in the ratio map of FUV to NUV intensity displayed in Figure \ref{fig_correlation_UV_NHI}(c) where the regions of the dust excesses (northeast and southwest quadrants) have ratio values close to 1. By construction of the plot of Figure \ref{fig_correlation_UV_NHI}c, a value of 1 indicates that the FUV and NUV maps correlate with a correlation coefficient compatible with the ones independently deduced from the correlation with \nhi. Because we subtracted the isotropic intensities at both wavelengths in the ratio, any faint component present at one wavelength and not in the other readily shows up in the ratio. This is why the faint NUV extra component (or zodiacal residual) drives the ratio map to blue values toward 0.5, but the regions of the dust excesses do not strongly deviate from 1.

\subsection{UV Extragalactic Background}

Measuring of the UV extragalactic background started 50 yr ago. \cite{2019arXiv190905325M}, Table 1, reviews 75 publications related to observations of diffuse UV among which 23 resulted in an assessment of the FUV extragalactic background with an average of 283 photons s$^{-1}$cm$^{-2}$sr$^{-1}$$\AA$$^{-1}$. Recently \cite{2019MNRAS.489.1120A} obtained an isotropic value of 354$\pm$18 photons s$^{-1}$cm$^{-2}$sr$^{-1}$$\AA$$^{-1}$ at FUV wavelength and 588$\pm$37 photons s$^{-1}$cm$^{-2}$sr$^{-1}$$\AA$$^{-1}$ at NUV wavelength  in a study of the region between the Galactic latitudes 70$^{\circ}$ and 80$^{\circ}$.

\cite{2018ApJ...858..101A} calculated the expected UV extragalactic background taking into account photons from galaxies, quasars, and intergalactic medium. They obtained a value of 114$\pm$18 photons s$^{-1}$cm$^{-2}$sr$^{-1}$$\AA$$^{-1}$ in the GALEX FUV band and 194$\pm$37 photons s$^{-1}$cm$^{-2}$sr$^{-1}$$\AA$$^{-1}$ in the GALEX NUV band. In \cite{2000ApJ...542L..79G}, an expected range of 144–195 photons s$^{-1}$cm$^{-2}$sr$^{-1}$$\AA$$^{-1}$ at FUV wavelength and 179 photons s$^{-1}$cm$^{-2}$sr$^{-1}$$\AA$$^{-1}$ at NUV wavelength were found from a count of resolved objects in the Hubble Deep Fields observations. Those predictions differ from the measures, and this discrepancy led to hypothesizing for alternative sources of diffuse UV emission  \citep{2021arXiv210709585K,2015ApJ...798...14H,2013ApJ...779..180H}.

It is possible that the discrepancy between predicted and measured values of the UV extragalactic background fluxes could in part be due to the scattering on the dust excess displayed in Figure \ref{fig_dcm} that was not accounted for in the calculation of the foreground Galactic emission.
In the scatter plots of Figures \ref{fig_correlation_UV_NHI}(a) and (b), the y-intercepts of the dashed lines represent the diffuse UV intensities extrapolated to \nhi=0. As mentioned earlier, we found a value of $I_{0(FUV)}$=137$\pm$15 photons s$^{-1}$cm$^{-2}$sr$^{-1}$$\AA$$^{-1}$ for the FUV y-intercept, which is compatible with the prediction of \cite{2018ApJ...858..101A}. However the NUV y-intercept $I_{0(NUV)}$=378$\pm$45 photons s$^{-1}$cm$^{-2}$sr$^{-1}$$\AA$$^{-1}$ is still about twice as much as the NUV extragalactic background  prediction.
{Given the discrepancy between the measured and the expected NUV extragalactic background, we separately fitted the eastern and western halves of the mask region. We obtained y-intercepts compatible within 1$\sigma$ with the value derived in the whole mask. Our measure of the extragalactic background is therefore not significantly influenced by possible high-latitude or zodiacal excesses 
}

Our simplistic model based on a direct comparison with \nhi assumes a completely isotropic scattering of UV photons on dust grains; still we think that our work represents an improvement compared to previous studies that used simple latitude cuts. We tested the model of \cite{1979ApJ...227..798J} where the dust scattered intensity is $I_{UV}=a I_{0}\tau_{V} (1-1.1g\sqrt{sin(|b|)})+ I_{iso}$ where $I_{0}$ relates to the constant plane source that illuminates the high-latitude dust, $I_{iso}$ is the isotropic contribution, $\tau_{V}$ is the optical depth, $a$ is the albedo, and $g$ is the forward scattering asymmetry factor of Galactic dust grains, but we were not able to clearly disentangle the contributions from $I_{iso}$, $a$ and $g$. A better characterization of the UV extragalactic background probably requires a Monte Carlo simulation similarly to the work of \cite{ 2016MNRAS.459.1710M} together with a careful 3D distribution of all the local dust including the excesses observed in our work.  

\section{Interpretation and Perspective}

\begin{figure*}
  \centering
  \includegraphics[width=\hsize]{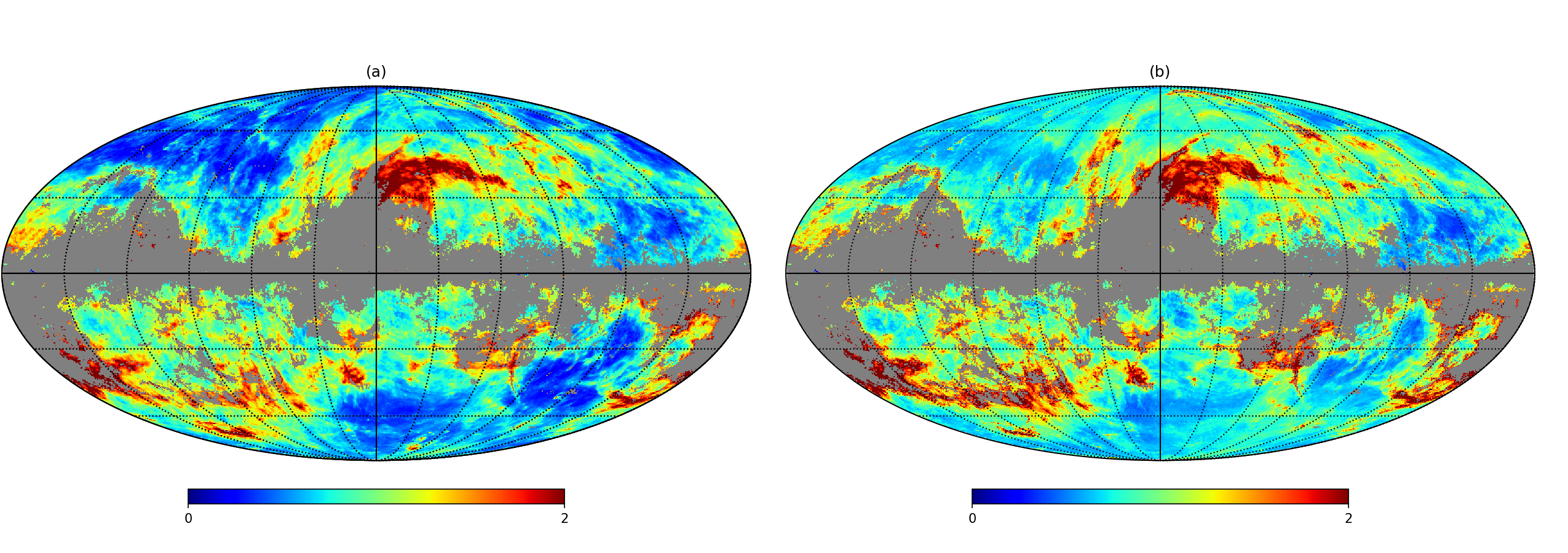}
  \caption{Spatial distribution of \nhi (a) and $\tau_{353}/\sigma_{e 353}$ with  $\sigma_{e 353}$=8.9$\times$10$^{-27}$ cm$^2$ (b) both divided by 2.4$\times$10$^{20}$ cm$^{-2}$/sin($|$b$|$). Maps are in Mollweide projection with a linear color scale and a 30$^{\circ}$ spaced grid. We masked pixels with
$\tau_{353}/\sigma_{e 353}-N_{\text{H}\,\textsc{i}}>2\times10^{20}$ cm$^{-2}$ 
where $\tau_{353}$ is not a good tracer for \nhi.}
  \label{cosec}
\end{figure*}

In this work, we observed, through their FIR emission and UV scattering, an excess of dust grains with a spatial extension covering 25\% of the sky. This excess of dust is anticorrelated with \nhi and more precisely with $N_{\text{H}\,\textsc{i}}\times$sin($|$b$|$) (Figure \ref{fig_anticorrel}). We can also see this anticorrelation in Figure \ref{cosec}a that represents \nhi divided by 2.4$\times$10$^{20}$ cm$^{-2}$/sin($|$b$|$) to emphasize the local gas distribution. In this map, the gas deficits in the northeast and southwest (dark blue holes) correspond to excesses in the maps of Figures \ref{fig_opacity_map} (top) and \ref{fig_FUV_opacity_map} (a and b).
The dust excesses noted in Figures \ref{fig_opacity_map} (top) and \ref{fig_FUV_opacity_map} (a and b) therefore seem to relate to local deficits in the WNM.
We further note that these deficits surround the inclined poles of the local ISM disk structure (see Figure 4 of \cite{1999A&A...346..785S}; \cite{2019A&A...625A.135L}).
Figure \ref{cosec}b displaying $\tau_{353}/\sigma_{e 353}$ divided by  2.4$\times$10$^{20}$ cm$^{-2}$/sin($|$b$|$) shows that the total gas potentially traced by $\tau_{353}/\sigma_{e 353}$ follows the cosecant trend (more uniform cyan and green colors toward these regions). \cite{2010ApJ...724.1389M} and \cite{2013ApJ...779..180H} noted a similar agreement for the GALEX UV intensity maps.

In our FIR study, the only way we found to strongly reduce this excess of dust was to set the initial 100 $\mu$m map zero level to values below -0.6 \Iunits thereby significantly increasing the 100 $\mu$m intensity for low \nhi regions. This increase would translate for the fit in a significant increase in $T$ and a decrease in \ttau for regions of low \nhi in better agreement with the prediction based on \nhi and the excess would cancel.
We disregarded this possibility for two reasons. First a value of -0.6 \Iunits is largely rejected by the two calibration methods illustrated by Figures \ref{fig_zero_100} and \ref{fig_zero_100_with_Dirbe}. Secondly that would result in an average temperature of 25~K for dust with $|$b$|>$60$^{\circ}$ while the average temperature for the rest of the sky would be 21~K. Since big grains are in thermal equilibrium with the ambient interstellar radiation field, their temperatures trace the local UV and optical radiation field intensity \citep{2010ApJ...724L..44C}, and the distribution of hot O and B stars that follows the Gould Belt \citep{1997A&A...323L..49P} could not be at the origin of a large increase of temperature toward the Galactic poles.

Dust particles obscure the light from faint and distant sources. Observations with cosmological purposes often use long exposure in deep fields carefully chosen for, among other criteria, their low dust column density. The Euclid mission \citep{2011arXiv1110.3193L} has for example chosen 3 deep fields\footnote{\url{https://www.cosmos.esa.int/web/euclid/euclid-survey}}: the Euclid Deep Fields North (EDFN), the Euclid Deep Fields South (EDFS), and the Euclid Deep Fields Formax (EDFF). We estimated that on average the dust column density is 22\% higher in the EDFN than predictions based on \ttau taken from Planck XLVIII, 74\% higher in EDFS, and 90\% higher in EDFF. This number reaches 120\% within 2$^{\circ}$.5 of the Lockman Hole.    

If some gas is associated to the dust seen in excess, it should interact with CR and produce $\gamma$ rays from $\pi^{0}$ decay and Bremsstrahlung. With an average column density of 0.7$\times 10^{20}$ cm$^{-2}$ over 25\% of the sky, we predict that more than 5$\times$10$^5$ counts above 50 MeV in 10 years of Fermi-LAT data should arise from those interactions. This gas, assuming the same  $\gamma$-ray emissivity as the local hydrogen, should be easily detectable by Fermi-LAT. It will be the subject of a forthcoming publication.

\section{Conclusions}

In this work, we compared, outside the Galactic plane, the spatial distribution of the dust traced by its FIR, NUV, and FUV emission to the spatial distribution of atomic hydrogen derived from radio surveys. The nonuniformity of the spatial distribution of the FIR opacity found in previous studies led to the investigation of a possible issue in the zero level of FIR maps.

We measured the zero level (detector offset plus CIB monopole) of the Planck-HFI skymap at 857 GHz in a region where we knew from the Fermi-LAT diffuse analysis that \nhi provides a good template for the interstellar $\gamma$-ray sky. From that value, we deduced the zero levels of Planck maps at 545 and 353 GHz as well as the zero level of the 100 $\mu$m map from COBE/DIRBE and IRAS and zero levels of COBE/DIRBE maps. We refined those values using the monopoles of the residuals between the intensity maps and an iteratively fitted MBB model. We obtained significantly different zero levels from those deduced in regions of lowest \nhi in Planck XI and in Planck XLVIII. The best fits of the MBB model FIR emission at 100 $\mu$m and 857, 545, and 353 GHz lead to significantly different maps of the dust opacity spectral index \bb and temperature $T$ and an overall increase in the optical depth \ttau of 7.1$\times$10$^{-7}$ compared to the results of Planck XLVIII.

We compared \ttau and \nhi and obtained an average dust opacity of $\sigma_{e 353}$=(8.9$\pm$0.1)$\times$10$^{-27}$ cm$^2$. From the average opacity, we deduced a dust-to-gas mass ratio of 0.53$\times$10$^{-2}$. The opacity displays a uniform spatial distribution outside regions with dense gas except in various large-scale regions of low \nhi ($\lesssim$2$\times 10^{20}$ cm$^{-2}$) that cover 25$\%$ of the sky. In those regions, we observe that on average the dust column density is 45\% higher than predictions based on \nhi. The excess reaches a maximum around the Lockman Hole where the region within 5$^{\circ}$ of this position contains 2.5 times more dust than what is expected from \nhi. The spatial distribution of this excess is not what one would expect from dust associated to a WIM-Reynolds-type layer. 

We calculated the rate of NUV and FUV photons detected by GALEX per atomic hydrogen and studied its spatial distribution. We observed an excess of photons with a similar spatial distribution to what we obtained in our FIR study. The dust excess is then both observed through its FIR and UV emission. This excess of dust can also be interpreted as a gas deficit in regions of low \nhi and more precisely in regions that depart from a cosecant law.

We measured with a simplistic modeling a FUV isotropic intensity of 137$\pm$15 photons s$^{-1}$cm$^{-2}$sr$^{-1}$$\AA$$^{-1}$ in agreement with extragalactic flux predictions and a NUV isotropic intensity of 378$\pm$45 photons s$^{-1}$cm$^{-2}$sr$^{-1}$$\AA$$^{-1}$ corresponding to twice larger than the predictions.

\vspace{5mm}

The skymaps of the MBB dust parameters \ttau, $T$, and \bb displayed in Figure 7 (left) are available for download in Fits format\footnote{\url{https://irfu.cea.fr/dap/en/Phocea/Vie_des_labos/Ast/ast_technique.php?id_ast=5060}}.

\begin{acknowledgements}
J.M.C. would like to thank Emanuele Daddi, Guillaume Laibe, Konstancja Satalecka, Marc Sauvage, and Vincent Tatischeff for helpful conversations; 
Jayant Murthy for kindly providing an updated version of his diffuse FUV Monte Carlo code\footnote{\url{https://zenodo.org/record/4496308}}; Julia Casandjian and Iroise Casandjian for their careful reading of the manuscript.
\end{acknowledgements}

%



\software{NumPy \citep{2011CSE....13b..22V}, Matplotlib \citep{2007CSE.....9...90H}, FILLTEX \citep{2017JOSS....2..222G}, HEALPix \citep{2005ApJ...622..759G}, ROOT \citep{1997NIMPA.389...81B}
}



\bibliography{mybiblio}{}
\bibliographystyle{aasjournal}



\end{document}